\documentclass[a4paper, 11pt]{article}
\usepackage{fullpage}
\usepackage[textwidth=18cm, textheight=26cm, centering]{geometry}
\usepackage{lineno}
\usepackage{gensymb}
\usepackage{rotating}
\usepackage{tabularx}
\usepackage{float}
\usepackage{textcomp}
\usepackage{booktabs}
\usepackage[version=4]{mhchem}
\usepackage{graphicx}
\usepackage{color}
\usepackage{siunitx}
\usepackage[superscript,biblabel]{cite}
\usepackage{authblk}
\usepackage{hyperref}
\usepackage{blindtext}
\usepackage{multirow}
\usepackage{booktabs} 
\usepackage{enumitem}
\usepackage{siunitx}
\let\DeclareUSUnit\DeclareSIUnit
\let\US\SI
\DeclareUSUnit\inch{in}
\DeclareSIUnit{\rpm}{rpm}
\DeclareSIUnit{\fps}{fps}
\DeclareSIUnit{\molar}{M}
\usepackage{chemformula}
\usepackage{import}
\usepackage{setspace}
\title{Integrated impedance sensing of liquid sample plug flow enables automated high throughput NMR spectroscopy}

\author[1]{Omar Nassar}
\author[1]{Mazin Jouda}
\author[1]{Michael Rapp}
\author[1]{Dario Mager}
\author[1]{Jan G.\ Korvink}
\author[1]{Neil MacKinnon*}

\affil[1]{Institute of Microstructure Technology, Karlsruhe Institute of Technology (KIT), Hermann-von-Helmholtz-Platz 1, 76344 Eggenstein-Leopoldshafen, Germany}

\date{}                  
\renewcommand\Affilfont{\itshape\small}

\begin{document}
\maketitle
\subsection*{*Corresponding author:}
Hermann-von-Helmholtz-Platz~1,\\
76344 Eggenstein-Leopoldshafen, Germany\\
Phone number: +49 721 608-22740\\
Fax number: +49 721 608-24331\\
Email address: neil.mackinnon@kit.edu

\vspace{0.5in}

\subsection*{Email addresses of all authors:}
\begin{table}[h]
    \begin{tabular}{l|l}
	Author & E-mail address \\
	\midrule
	Omar Nassar & omar.nassar@kit.edu \\
	Mazin Jouda & mazin.jouda@kit.edu \\
	Michael Rapp & michael.rapp@kit.edu \\
	Dario Mager & dario.mager@kit.edu \\
	Jan G. Korvink & jan.korvink@kit.edu \\
	Neil MacKinnon & neil.mackinnon@kit.edu 
	\end{tabular}
\end{table}
	
\vspace{0.5in}

\newpage
	
\section*{Abstract}
A novel approach for automated high throughput NMR spectroscopy with improved mass-sensitivity is accomplished by integrating microfluidic technologies and micro-NMR resonators. A flow system is utilized to transport a sample of interest from outside the NMR magnet through the NMR detector, circumventing the relatively vast dead volume in the supplying tube by loading a series of individual sample plugs separated by an immiscible fluid. This dual-phase flow demands a real-time robust sensing system to track the sample position and velocities and synchronize the NMR acquisition. In this contribution, we describe an NMR probe head that possesses a microfluidic system featuring: i) a micro saddle coil for NMR spectroscopy and ii) a pair of interdigitated capacitive sensors flanking the NMR detector for continuous position and velocity monitoring of the plugs with respect to the NMR detector. The system was successfully tested for automating flow-based measurement in a \SI{500}{\mega\hertz} NMR system, enabling high resolution spectroscopy and NMR sensitivity of \SI{2.18}{\nano\mole \ \second^{1/2}} with the flow sensors in operation. The flow sensors featured sensitivity to an absolute difference of 0.2 in relative permittivity, enabling distinction between most common solvents. It was demonstrated that a fully automated NMR measurement of nine individual \SI{120}{\micro\liter} samples could be done within \SI{3.6}{\min} or effectively \SI{15.3}{\second} per sample.

\vspace{1.5in}
\noindent
{\textbf{Keywords:} Microsystems, Microfabrication, Microfluids, High Throughput, Probe Design, Impedance Sensor, Automation}
\newpage
	
\section{Introduction}
Nuclear magnetic resonance (NMR) spectroscopy is an information dense analytical technique; however, current methodology is primarily limited to single sample measurements under low-throughput conditions.  There is therefore a disconnect between applications requiring screening of large parameter spaces, and the desire to take advantage of this information-rich technique. By addressing sample throughput, NMR could then be routinely applied to diverse fields including chemical process monitoring, drug discovery screening, and clinical analysis.

Several approaches have been developed to transition to high-throughput NMR analysis. Commercial automated sample changers are available, capable of queuing up to $\sim$ 500 samples for measurement~\cite{lilly1996sample, ross2001automation}. Here each sample is physically separated and thus cross-contamination is avoided, with the disadvantage that after each sample exchange the NMR instrument must be re-adjusted (tuning, matching, shimming). Instead of exchanging individual sample containers, a flow-through NMR system can be used to deliver a series of successive samples, separated by an inert gas, from outside the magnet through the NMR detector~\cite{gokay2012single}. This approach has proven to be a reliable alternative to enhance sample throughput, with additional enhancement enabled by designing probes with several radiofrequency (RF) NMR detectors~\cite{macnamara1999multiplex, wolters2002nmr, macnaughtan2003high, wang2004eight}. A flow-based approach is extremely attractive when sample quantities are small; instead of diluting the sample or dealing with poor filling factors using standard NMR hardware, microfluidic and micro-NMR techniques can be used. If the microfluidic sample handling is robust, the approach is fully compatible with standard sample quantities and thus is broadly applicable. 

Microfluidic-compatible NMR has continued to evolve since the first micro-NMR detector reports. In high-field NMR spectroscopy ($^1$H resonant frequency $>$ 200 MHz), this has included using microfluidic systems, often implementing flow, for chemical reaction monitoring~\cite{Zhivonitko_Characterization, Brcher_Thermostatted_2014, Brcher_Application_2016, Tijssen_Monitoring_2019,Swyer_Digital_2019}, biological sample monitoring~\cite{Kalfe_Looking, Yilmaz2016, Montinaro_3D_2018}, sensitivity enhancement strategies~\cite{ Mompen_Pushing, Carret_Inductive_2018, Eills_High_2019, C8LC01259H}, and recently featuring electrochemical functionality~\cite{Davoodi_An_2020}. Except for the digital microfluidic approach~\cite{Swyer_Interfacing_2016,Swyer_Digital_2019}, the sample in these examples has always been drawn from a reservoir without explicitly considering the situation of handling \textit{individual} samples. Sample loading should be done from outside of the magnet, and thus there is the challenge of reducing dead-volumes in filling transfer lines (often a few meters in length) before actually reaching the NMR detection volume. To achieve high sample throughput while taking advantage of microfluidic flexibility, it would be advantageous to convert the dead volume into a multi-sample queue (as in the automated sample exchange systems) which can be accomplished using two-phase flow. Under this condition, tracking sample positions must be addressed to ensure each NMR spectral acquisition is synchronized to the correct sample.

Fluids are commonly detected in microfluidic applications by integrating either optical~\cite{nguyen2006optical, tkaczyk2011microfluidic} or electrical sensing systems~\cite{elbuken2011detection}. Electrical sensing is attractive for two-phase flow because i) samples are often optically similar, and ii) immiscible liquids with significant differences in dielectric constants (relative permeability) are readily available. Aqueous samples will have a dielectric constant at least one order of magnitude larger than a typical immiscible oil, making electrical sensing an ideal candidate for plug detection. Sensing fluids based on their electrical properties can be achieved using capacitive or resistive sensors; resistive sensors require direct contact between the target fluid and sensing electrodes and thus could interfere with the sample (e.g. influence temperature) and/or become a source or cross-contamination~\cite{srivastava2006electronic, cole2009multiplexed}. On the other hand, capacitive sensors can detect fluids without any contact between the target fluid and the electrode surface~\cite{elbuken2011detection, soenksen2018closed}. Capacitive sensors have been utilized in a variety of microfluidic applications, including detecting the speed and volume of microdroplets~\cite{elbuken2011detection}, fluid-level sensing~\cite{soenksen2018closed, wells1991capacitive}, estimation of liquid composition~\cite{ghafar2007hybrid} and in pressure measurements~\cite{shih2003surface}.

This study presents a novel approach combining  microfluidic technologies with micro-NMR detectors to realize fully automated high-throughput NMR spectroscopy with high mass-sensitivity. A scalable, NMR compatible, low power, and cost-effective microfluidic system is developed that has a micro-saddle NMR detector rolled around a glass capillary flanked by two integrated impedance sensors for sample position and velocity measurement. The system operates by loading plugs of target samples separated by an immiscible fluid into a flexible microfluidic tube. This train of samples is kept under flow and fed through the capillary with the micro detector and sensors. The impedance sensors detect the interface between aqueous and oil plugs, which a microcontroller then uses to synchronize the NMR spectroscopy acquisition regardless of the flow velocities and sample volume. This technology operates in a plug-based approach where each plug could be a different sample, reducing the required sample volume for flow NMR experiments (microliter sample volumes are demonstrated). The continuous flow regime enabled by the system improves the experimental time per sample considerably by continuously supplying freshly polarized sample. The design, optimization, and validation of the flow sensing micro-NMR system are discussed. A robust micro-fabrication process is introduced for producing the micro-coil and sensors. The system's signal-to-noise ratio (SNR) was characterized as a function of the sample volume, flow rate, and analyte concentration. Additionally, a proof-of-concept analysis of various commercial beverages is presented as evidence that the system can efficiently automate NMR spectroscopy on different samples.

\section{Results}
\begin{figure*}[t!]
\centering
\includegraphics[width=\textwidth]{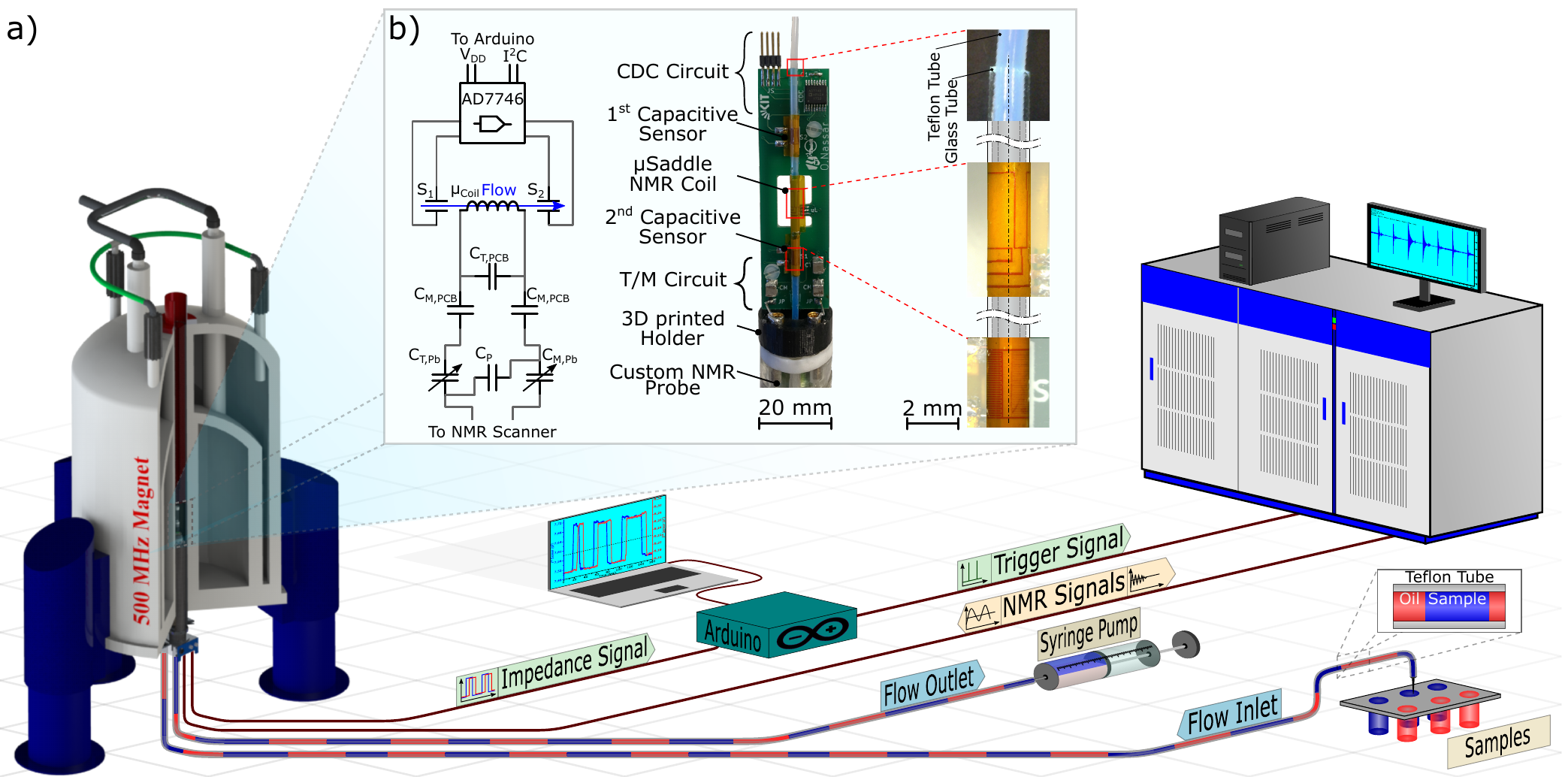}
\caption{(a) An overview schematic of the main components of the new high throughput NMR experimental setup. A series of plugs of different samples separated by an immiscible fluid are hydraulically  transported in a flow tube (OD/ID, \SI{1.6/1}{\milli\meter}, Teflon) through a custom-built NMR probe mounted in an \SI{11.7}{\tesla} NMR magnet. The flow is controlled from outside the magnet using a standard syringe pump. A microcontroller is used to drive capacitive sensors on the probe head to sense the position and the velocity of the samples and synchronise the NMR acquisition. (b) Photograph of a custom-built probe head mounted on the NMR probe. The probe head has a PCB that possesses a micro NMR detector with pre-tuning and matching network, two capacitive sensors with their interface electronics, and a glass capillary through which the sample flow tube is fed. A 3D printed holder supports the PCB. Right: a microscopic image for the NMR detector, the capacitive sensor, and the insertions of the PTFE tube inside the glass capillary. Left: schematic of the equivalent electrical circuit in the NMR probe.
}
\label{fig:full_setup}
\end{figure*}

\subsection{System description}
In the following section, we introduce the technical details of the flow-sensing system for actively triggering NMR acquisition and present the results of the flow NMR experiments using this automation system. An overall schematic of the actively triggered NMR system hardware components can be seen in Fig.~\ref{fig:full_setup}a, with a detailed figure of the NMR probe head in Fig.~\ref{fig:full_setup}b. The custom-built NMR probe head consists of three main modules on a printed circuit board (PCB): 
(I) a micro NMR detector along with pre-tuning and matching capacitors; (II) two capacitive sensors, flanking the NMR detector, connected to a capacitance-to-digital converter (CDC) microchip to transfer the capacitance reading outside the magnet; (III) a glass capillary to contain the microfluidic flow. The PCB, with these three modules, is supported by a 3D printed holder mounted on a custom NMR probe. 

\textit{Module (I).} The probe head was designed so that, when inserted into the NMR static magnet field ($B_0$), the NMR micro-detector is located in the magnet's high homogeneity region (isocenter). This also guarantees that the detection volume is in the center of the spectrometer shim system, used to correct minor $B_0$ field deviations. The probe head was designed to be compatible with a triple-axis gradient sleeve (Bruker), so that the system is additionally compatible with magnetic resonance imaging (MRI) experiments and pulsed field gradient (PFG) sequences. To reduce material interfaces and sources of background signal in the vicinity of the NMR detector, the PLA of the holder and FR4 of the PCB were removed next to the NMR detector. The pre-tuning and matching network on the PCB enables straightforward adjustment of the micro-detector resonant frequency, so that NMR experiments could, in principle, be done on any NMR sensitive nucleus.  The probe head then becomes a plug-and-play unit, with the NMR probe trimmers used only to fine-adjust the resonant circuit. In addition to standard externally controllable tuning and matching trimmers, the custom NMR probe featured an inlet and an outlet port for the microfluidic tubing and a channel for the CDC's electrical connection.

\textit{Module (II).} Two capacitive sensors were placed before and after the NMR detector with a defined separation of \SI{10}{\milli\meter} w.r.t.\ the NMR detector. Using a sensor before and after the NMR detector enables velocity calculation, since the total separation is specified (\SI{30}{\milli\meter}). Furthermore, the triggering experiment is independent of the flow direction. The capacitive sensors were connected to an Arduino microcontroller unit positioned outside the NMR magnet through the probe head's CDC microchip. The microcontroller was connected to the triggering signal input-output unit (TTL I/O) of the NMR spectrometer console and was used to synchronize NMR data acquisition.

\textit{Module (III).} A microfluidic tube (OD/ID, \SI{1.6/1}{\milli\meter}) was fed from outside of the magnet through the custom NMR probe, and a through a glass capillary (OD/ID, \SI{2.15/1.72}{\milli\meter}) on the probe head on which the NMR micro-detector and flow sensors were mounted. There are two advantages in using this configuration: (I) on the microfluidic side, it avoids fluidic junctions, preventing sample leakage or flow disturbance which often creates air bubbles; (II) on the NMR side, the sample is then confined to the homogeneous excitation $B_1$ field volume of the micro-NMR detector (field distortion occurs near the coil tracks). Sample flow was enabled using a syringe pump located outside of the NMR magnet.

\subsection{NMR detector - design and characterization}
\begin{figure*}[t!]
\centering
\includegraphics[width=0.88\textwidth]{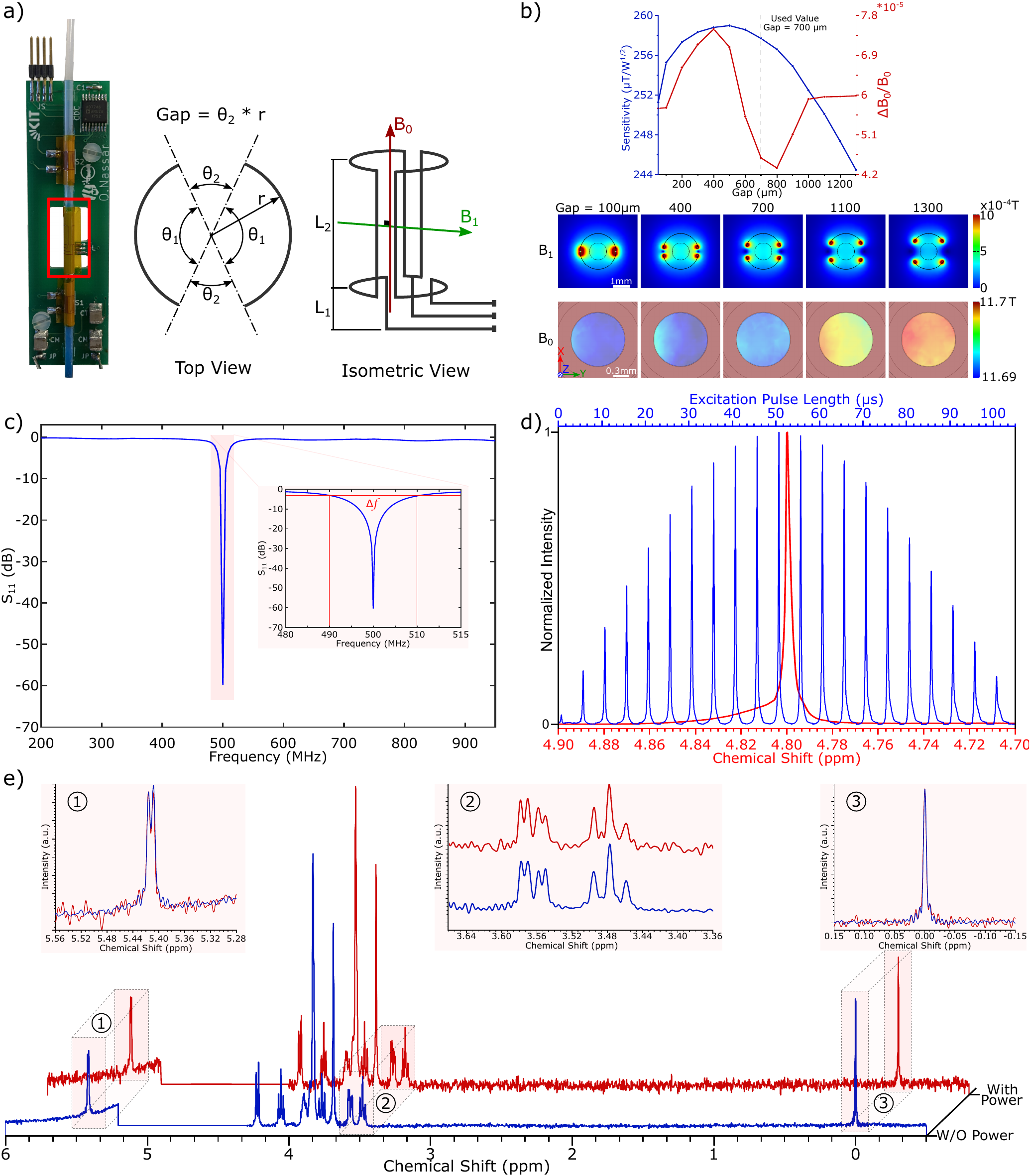}
\caption{(a) Photograph (left) and schematics of a saddle coil. Design parameters include the gap between the two conductor loops, the length of the coil (L\textsubscript{2}), and the length of the current leads (L\textsubscript{1}). (b) The simulated coil sensitivity and the B\textsubscript{0} homogeneity versus the gap (top). The B\textsubscript{0} and B\textsubscript{1} field profiles at a cross-section in the center of the coil are plotted for different gap values (bottom). (c) The reflection coefficient S\textsubscript{11} of a micro saddle coil connected to the custom-built probe, tuned and matched to \SI{500}{\mega\hertz} and \SI{50}{\ohm}. The resonance width at \SI{-3}{\decibel} was measured to be approximately \SI{20}{\mega\hertz}, resulting in a quality factor of 50. (d) Excitation pulse nutation spectrum (blue, upper scale) and NMR spectrum (red, lower scale) of water. Measurements were done using a micro saddle coil at \SI{500}{\mega\hertz}. The delay between consecutive nutation measurements was \SI{10}{\second}. (e) Comparison of single-scan $^1$H NMR spectra before (blue) and after (red) connecting power to the capacitive sensing electronics. The sample was stationary in this experiment (\SI{250}{\milli\molar} sucrose and \SI{30}{\milli\molar} TSP in DI water). SNR was observed to decrease by $\sim$\SI{55}{\percent}. The regions between 5.56 - 5.28 ppm, 3.68 - 3.36 ppm, and 0.15 - -0.15 ppm are re-plotted in red boxes. The intensity of the water peak (5.2 - 4.3ppm) was set manually to zero.}
\label{fig:microcoil}
\end{figure*}

Key design considerations for the RF detector to maximize performance (spectral resolution, sensitivity) are: (i) the influence of the detector conductive tracks on the homogeneity of the primary static magnetic field B\textsubscript{0}; (ii) the homogeneity and strength of the excitation magnetic field B\textsubscript{1} generated by the coil. The homogeneity of B\textsubscript{0} can vary due to a magnetic susceptibility mismatch between material interfaces within the RF region, directly impacting the spectral resolution. The strength and homogeneity of the B\textsubscript{1} field in the sample volume affects the measurement sensitivity. To optimize the geometrical dimensions of the saddle coil, finite element simulations were performed. The optimization aimed to determine the coil geometry in a way to increase the generated RF field $B\textsubscript{1}$ per unit current while keeping the $B\textsubscript{0}$ and $B\textsubscript{1}$ fields as homogeneous as possible. The simulation results are presented in Fig.~\ref{fig:microcoil}b, the coil sensitivity and the B\textsubscript{0} field homogeneity are plotted against the gap (arc distance between the loops). Furthermore, the distribution of the B\textsubscript{0} field and B\textsubscript{1} field are presented at different gaps. The simulation results reveal that conductive tracks with a high degree of symmetry not only increase the sensitivity but also enhance the homogeneity of the B\textsubscript{0} and B\textsubscript{1} fields. Based on the simulation study, a gap of \SI{700}{\micro\meter} was selected for the coil fabrication (see Fig.~S1 for the simulated $B\textsubscript{1}$ field direction). The fabricated coil had a track width of \SI{100}{\micro\meter} with L\textsubscript{1} and L\textsubscript{2} of \SI{2}{\milli\meter} and \SI{4}{\milli\meter}, respectively (see Fig.~\ref{fig:microcoil}a for description). Fabrication details can be found in Materials and Methods.

The fabricated micro-NMR detector was evaluated using NMR spectroscopy measurements performed using an AVANCE III \SI{11.7}{\tesla} NMR system ($^1$H resonance frequency \SI{500}{\mega\hertz}, Bruker BioSpin) under stationary (non-flowing) conditions. The coil was filled with a sample (\SI{30}{\milli\molar} TSP dissolved in deionized (DI) water) using a tube with inner diameter \SI{1.0}{\milli\meter}, yielding an active volume of \SI{3.14}{\micro\liter}. The loaded quality factor was measured to be $Q = 50 \pm 2$ (Fig.~\ref{fig:microcoil}c). Spectral resolution was determined using the FWHM of the $^1$H NMR water resonance (Fig.~\ref{fig:microcoil}d, red spectrum), with the microsystem yielding a linewidth of \SI{1.25}{\hertz}. The excitation field performance was evaluated by measuring a nutation spectrum - a series of $^1$H NMR spectra at fixed applied power (\SI{1}{\watt}), varying the excitation pulse time from 1-\SI{105}{\micro\second}, incremented by \SI{5}{\micro\second} (Fig.~\ref{fig:microcoil}d, blue spectra). The time period of the nutation spectrum envelop was measured to be $T\textsubscript{nut}=\SI{204}{\micro\second}$ (\SI{90}{\degree} flip angle achieved at \SI{51}{\micro\second}), thus the nutation frequency was calculated to be $f\textsubscript{nut}=\SI{4.9}{\kilo\hertz}$. Using Equation~\ref{eq:b1exp}~\cite{levitt2013spin}, the B\textsubscript{1} field strength was calculated to be $B\textsubscript{1} \approx \SI{230}{\micro\tesla}$ ($^1$H gyromagnetic ratio $\gamma\textsubscript{\textsuperscript{1}H} = \SI{42.756}{\mega\hertz\per\tesla}$, angle between the B\textsubscript{1} and B\textsubscript{0} fields $\theta$ = \SI{90}{\degree}). 

\begin{equation}
    f\textsubscript{nut}=\frac{1}{2} \gamma B\textsubscript{1} sin(\theta).
    \label{eq:b1exp}
\end{equation}

 \noindent Thus, the micro-coil sensitivity ($\eta\textsubscript{P}$) was calculated using Equation~\ref{eq:coilsensitivity} to approximately $\SI{230}{\micro\tesla\per\sqrt{\watt}}$, comparable to the simulation result in Fig.~\ref{fig:microcoil}b.

To implement flow sensing, additional electronic components were added to the NMR system. These components add additional noise to the NMR signal, as demonstrated in Fig.~\ref{fig:microcoil}e.  A single scan $^1$H NMR experiment was done on a stationary sample (\SI{250}{\milli\molar} sucrose and \SI{30}{\milli\molar} TSP in DI water) before and after connecting the periphery sensing components, resulting in an SNR reduction by 55\%, so that the nLOD\textsubscript{$\omega$} increases from \SI{1.19}{\nano\mole \ \second^{1/2}} to \SI{2.18}{\nano\mole \ \second^{1/2}}. Nevertheless, the NMR sensitivity was comparable to other micro-coils, even with the additional noise~\cite{badilita2012microscale}.

\subsection{Capacitive sensors - design and characterization}
\begin{figure*}[t!]
\centering
\includegraphics[width=0.83\textwidth]{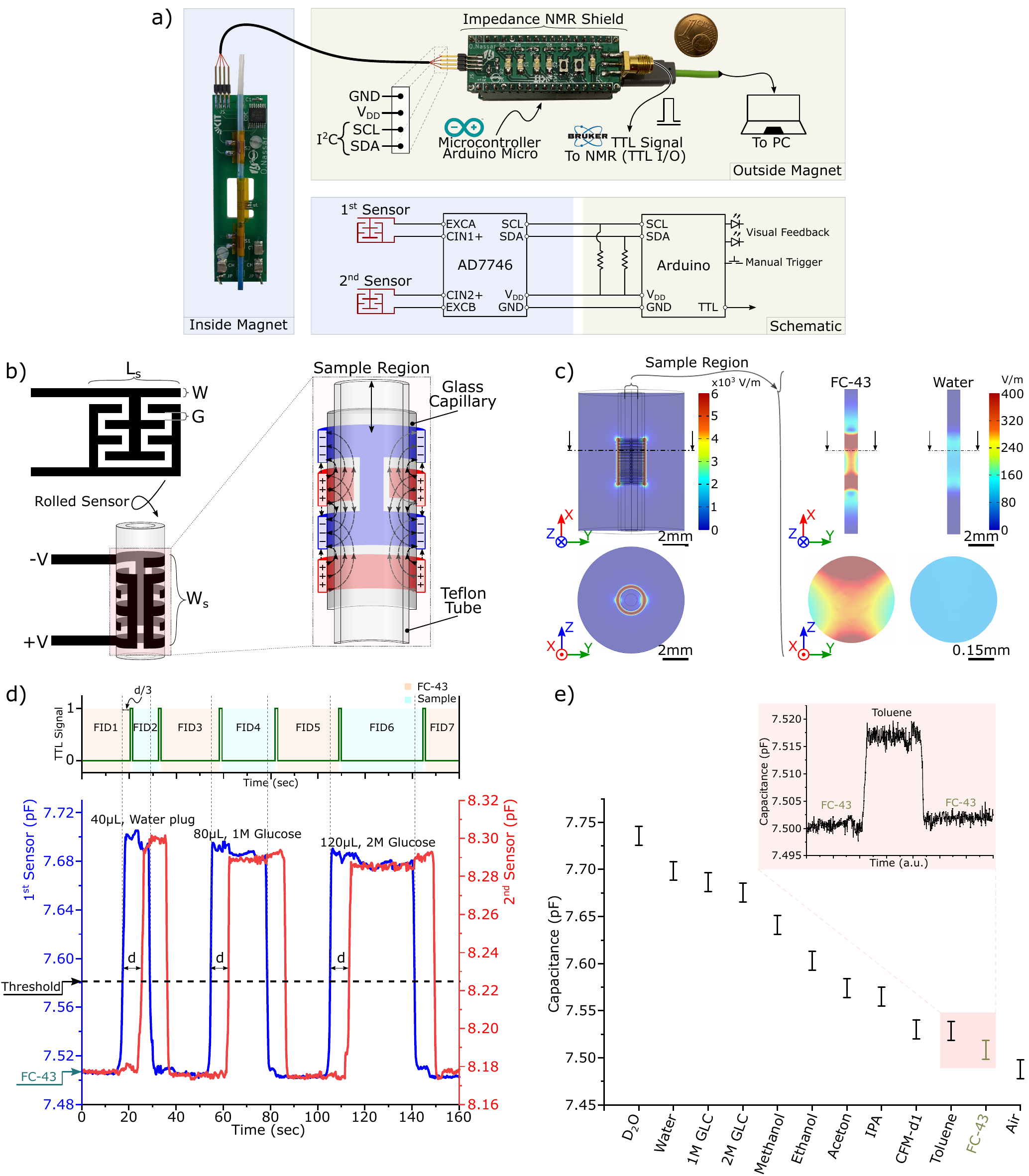}
\caption{(a) The electronic hardware components enabling the triggered NMR experiment. The capacitive sensors are connected to a CDC microchip on the probe head. The digitized output readings are transferred via an I\textsuperscript{2}C communication to the MCU outside the magnet. An Arduino shield is designed for the I\textsuperscript{2}C connections with LEDs for visual feedback, and push-button for manual triggering. 
(b) 2D schematic layout of the capacitive sensors and the expected fringe field after rolling around the glass capillary. 
(c) Finite element simulation (COMSOL Multiphysics) demonstrating the electric field distribution inside the capacitive sensor (water and FC-43 as examples). Left: an overview of the sensor, right: focusing on the target sensing region.
(d) The flow sensor output during a triggered NMR experiment. The experiment was done on three plugs of different volumes and different concentrations separated by plugs of FC-43 (\SI{80}{\micro\liter}). The flow velocity was \SI{4}{\milli\meter\per\second}. The TTL signals are sent from the MCU to the TTL I/O unit of the NMR console after a calculated delay, $d$/3. This ensures that a new NMR acquisition begins when the sample enters the NMR detection volume. The time $d$ is calculated based on the sample velocity, which is measured using the capacitive sensors (top). 
(e) A plot of measured capacitance for different solutions. The error in the measurement arises from the noise of the CDC chip ($\pm$~\SI{0.03}{\pico\farad}). The system is capable of differentiating toluene ($\varepsilon\textsubscript{r}=2.38$) from FC-43 ($\varepsilon\textsubscript{r}=1.9$) (inset).}
\label{fig:sensor}
\end{figure*}

Interdigitated electrodes were used for the detection of flow, presented in Fig.~\ref{fig:sensor}a. Interdigitated capacitive sensors (IDC) are sensitive to differences in material dielectric constants by modulating the sensor capacitance via the fringe field generated by the electrodes~\cite{mamishev2004interdigital}. Figure~\ref{fig:sensor}b illustrates a 2D layout of the sensor and the expected fringe field lines of the rolled sensor within the target region. Rolling the sensor around the capillary directs the fringe field towards the target fluid, which is more compact than traditional co-planar capacitive sensors and enables increased sensitivity to differences in permittivity.

To optimize the IDC performance, the width of the sensor, the total number of electrodes, and the gap between sensing electrodes were explored (Fig.~\ref{fig:sensor}b). The overall width of the capacitive sensor W\textsubscript{S} should be minimized to reduce the settling time of the sensor. Increasing the number of electrodes improves the smoothness of the signal~\cite{elbuken2011detection}. Additionally, the width of the gap G between the sensor electrodes affects the penetration depth of the fringe field as in the case of co-planar electrode capacitive sensors~\cite{li2006design}. Several design iterations were examined: a gap of G~=~\SI{25}{\micro\meter} was sufficiently small for the rolled IDC to sense the target fluids. The number of electrodes was N~=~30, each with a width of W~=~\SI{100}{\micro\meter}, resulting in a sensor with a total width of W\textsubscript{S}~=~\SI{3.725}{\milli\meter} and a length of L\textsubscript{S}=\SI{6.6}{\milli\meter} (slightly less than the capillary circumference \SI{6.75}{\milli\meter}).

The capacitance of the IDC was read-out using an off-the-shelf capacitance-to-digital converter (CDC) (AD7746, Analog Devices). The AD7746 microchip provides two separate measurement channels each of 24-bit resolution. The full-scale capacitance of the CDC is $\pm\SI{4}{\pico\farad}$ from an internal set value (\SI{0}{}-\SI{17}{\pico\farad}), with an accuracy of \SI{4}{\femto\farad} and a resolution down to \SI{4}{\atto\farad}. It also provides two built-in excitation sources (up to \SI{5}{\volt} at \SI{16} or \SI{32}{\kilo\hertz}) and filtering/amplifying circuits. The CDC has a 16-lead TSSOP package configuration (\SI{4.5} \times \SI{5}{\milli\meter}), which enabled integration on the NMR detector PCB. The update rate of the CDC can be set between \SI{10}-\SI{90}{\hertz} resulting in RMS noise levels of \SI{4.2}-\SI{40}{\atto\farad}. Each capacitive flow sensor was connected to a separate measurement channel of the CDC, via a positive input pin (CIN 1/2 (+)) and an excitation pin (EXC A/B) leveraging the single-end measurement. Excitation was set to \SI{5}{\volt} at \SI{32}{\kilo\hertz}, and the sampling rate was set to \SI{10}{\hertz}. The tracks connecting the capacitive sensors to the CDC were kept as far as possible from each other to prevent cross-coupling between the two sensors. The digitized capacitance measurements of the CDC were transmitted to a microcontroller unit outside the magnet via I\textsuperscript{2}C serial communication, as shown in Fig.~\ref{fig:sensor}a. An Arduino shield was developed for connecting the I\textsuperscript{2}C wire; it additionally featured LEDs for visual feedback and push-buttons for manually triggering the NMR acquisition.  Two Arduino shields have been developed compatible with Arduino Uno and Micro (Fig.~S2).

The functionality of the system to detect permittivity differences was validated by measuring the capacitance in the presence of various solvents, summarized in Fig.~\ref{fig:sensor}e. The system was able to differentiate between toluene ($\varepsilon\textsubscript{r}=2.38$) and FC-43 ($\varepsilon\textsubscript{r}=1.9$). A difference in capacitance was also observed between air and FC-43, opening the possibility of air-bubble detection. The sensitivity was independent of magnetic field, with only a constant offset of +\SI{0.3}{\pico\farad} observed when the sensor was in the magnetic field. The sensitivity to a difference in relative permittivity was determined to be $\Delta\epsilon_r= 0.2$.

The design of our system features two sensors, with the objective of enabling direct velocity measurement and bi-directional flow. The accuracy of the flow sensors to measure velocity was characterized using plugs of colored aqueous samples separated with FC-43 and measuring the flow using a camera (\SI{60}{\fps}). The plug velocities were extracted from the video data using motion analysis software (Kinovea), which were then compared to the velocities calculated by the flow sensing system. The velocity measurements were within 4\% of the camera results (measured outside of the magnet to facilitate video recording).

The capability of the sensors to synchronize NMR data acquisition was validated by loading the sample tube with aqueous sample plugs separated by FC-43, and the results are summarized in Fig.~\ref{fig:sensor}d. Three sample plugs of different volumes and concentrations of glucose were transported at \SI{4}{\milli\meter\per\second}. The results reveal that the aqueous solutions separated by FC-43 plugs generate a capacity change slightly above \SI{100}{\femto\farad}. Therefore, the signal-to-noise ratio (SNR) of the capacitive sensor is satisfactory even when the sensor is operated at the maximum sampling rate of \SI{90}{\hertz} (RMS noise of \SI{40}{\atto\farad}). As can be observed, the capacitance measurement was not identical between the two sensors, most likely originating from fabrication tolerances and the difference in proximity between each sensor and the CDC. Nevertheless, the working principle is unaffected since individual threshold values could be set for each sensor. Importantly, the signals are sufficiently distinct that a robust triggering algorithm could be developed, described in detail in the Materials and methods section.

\subsection{NMR performance - varying sample volume and flow rate}

\begin{figure*}[t!]
\centering
\includegraphics[width=0.9\textwidth]{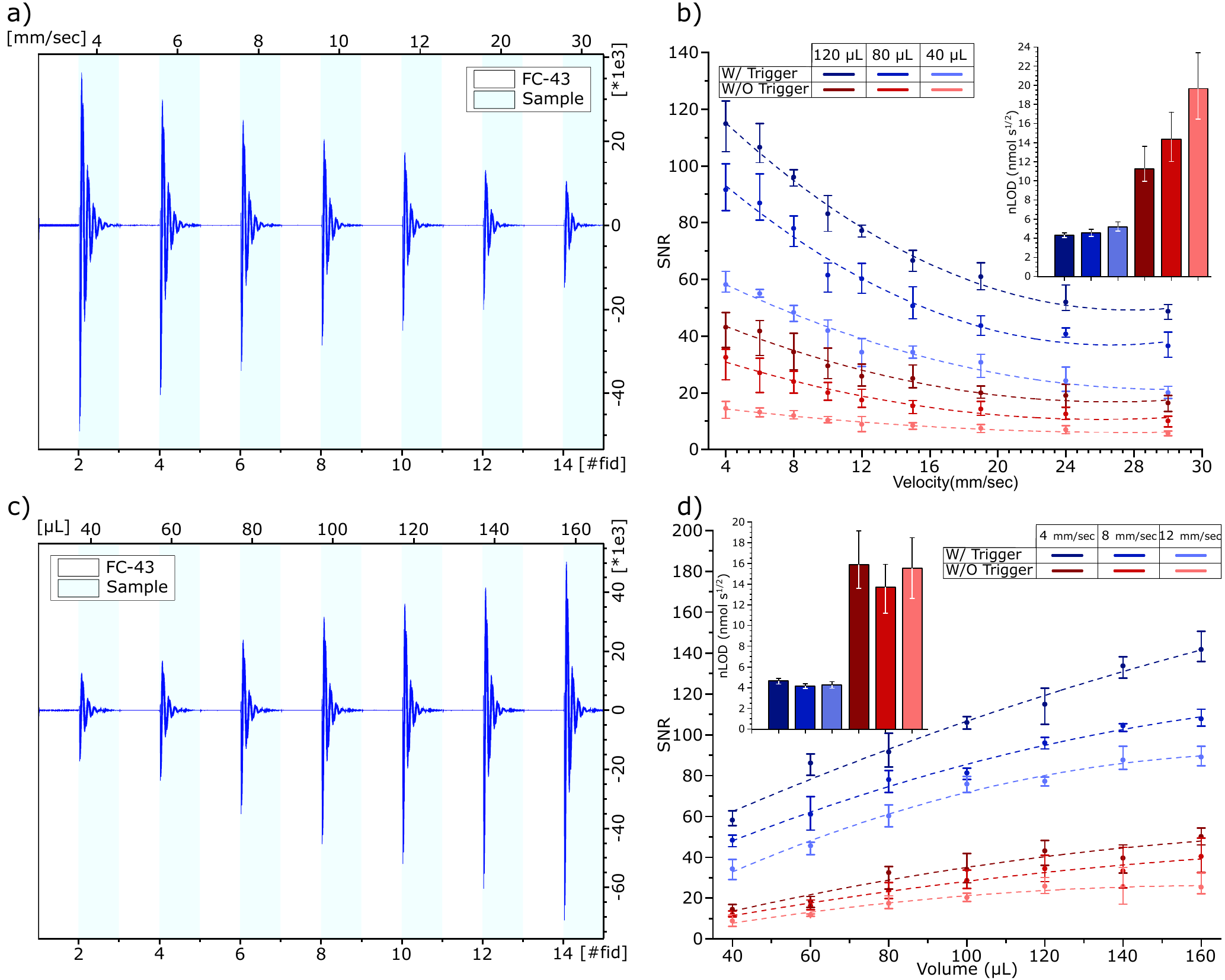}
\caption{(a) Measured FID from seven FC-43 oil and sample plugs, varying the flow rate. The sample volume was \SI{80}{\micro\liter}. (b) The relationship between TSP SNR (0 ppm) and the sample velocity as measured using the dual capacitive sensor. The sub-figure presents the nLOD\textsubscript{$\omega$} for each volume (mean $\pm$ standard deviation over all tested velocities). (c) Measured FID of seven oil and sample plugs varying the sample volume (oil volume held constant at \SI{80}{\micro\liter}). The flow velocity was \SI{8}{\milli\meter\per\second}. (d) The relationship between TSP SNR (0 ppm) and the sample volume. The sub-figure presents the nLOD\textsubscript{$\omega$} at each velocity (mean $\pm$ standard deviation over all tested volumes). In (b) and (d), each data point is mean $\pm$ standard deviation of the experiment performed in quadruplicate. The data were fit to a 2\textsuperscript{nd} order polynomial using the damped least-squares method.
}
\label{fig:flowandvolume}
\end{figure*}

NMR performance was characterized by monitoring the measured signal-to-noise ratio (SNR) using automatic and manual triggering of NMR data acquisition. In one series of experiments, the sample volume was held constant and the velocity was varied (\SI{4}{\milli\meter\per\second} to \SI{30}{\milli\meter\per\second
}). A series of sample volumes was tested (120, 80, and \SI{40}{\micro\liter}): using automatic triggering, all three sample plugs were loaded into the tubing, separated by \SI{80}{\micro\liter} of FC-43, and NMR acquisition was automatically started and stopped using the capacitive detection of the oil---water interface. For manual triggering, a single sample plug was loaded into the tubing, bordered by FC-43, and NMR acquisition was initiated and stopped when the sample entered and exited the probe.  In a second series of experiments, the sample volume was varied (\SI{40}{\micro\liter} to \SI{160}{\micro\liter}, incremented by \SI{20}{\micro\liter}) under constant flow rate. A series of flow rates was tested (4, 8, and \SI{12}{\milli\meter\per\second}): using automatic triggering, all sample plugs (separated by \SI{80}{\micro\liter} of FC-43) were loaded into the tube, while for manual triggering a single sample plug bordered by FC-43 was loaded. NMR acquisition was controlled as for the first series of experiments.  As a metric of performance, the SNR of the TSP signal (0 ppm) was determined under each test condition. The normalized limit of detection~\cite{badilita2012microscale} (nLOD\textsubscript{$\omega$}) was then calculated for each case based on the SNR of the TSP signal. The acquisition time for a single scan experiment was set to be \SI{0.299}{\sec} with an NMR sensitive volume of \SI{3.14}{\micro\liter}. The sample solution consisted of \SI{300}{\milli\molar} glucose, \SI{75}{\gram\per\liter} coloring powder, and \SI{30}{\milli\molar} TSP dissolved in DI water. 

\begin{figure*}[t!]
\centering
\includegraphics[width=\textwidth]{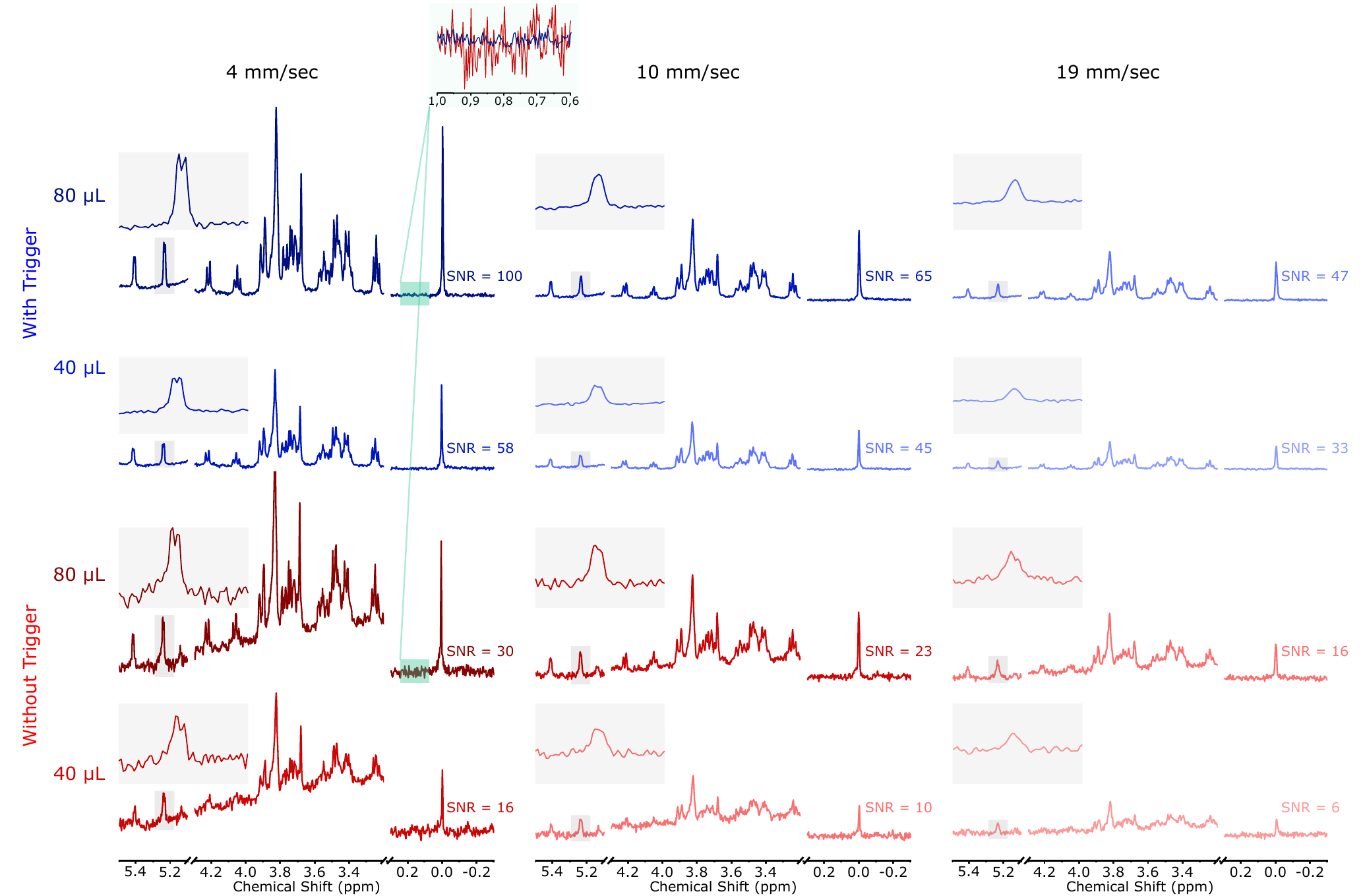}
\caption{
Selected $^1$H NMR spectra from the flow NMR spectroscopy experiments varying the sample plug volume and the flow rate (all extracted data plotted in Fig.~\ref{fig:flowandvolume}). The SNR of TSP for each spectrum is noted next to the signal (0 ppm). An expanded view of the glucose anomeric proton (5.2 ppm) is positioned next to each spectrum for spectral resolution comparison. The noise contribution to the signal is significantly reduced when the automated triggering system is implemented (inset, \SI{4}{\milli\meter\per\second}). The spectral region between 5.1-4.3 ppm (water resonance) was eliminated for clarity. The strong baseline distortion in the spectra without the automatic triggering system arises from signal accumulation from the oil phase. A full spectrum comparing auto- and manual triggering, and pure FC-43 samples are presented in Fig.~S5.
}
\label{fig:spectrum}
\end{figure*}

Figure~\ref{fig:flowandvolume}a and b report the results of the first series of experiments, where the flow rate was varied. Using the triggering system, the time-domain NMR signal (free induction decay, FID) for each oil and sample plug are measured and stored separately, as shown in Fig.~\ref{fig:flowandvolume}a. After the NMR measurement for an oil-sample plug pair was complete, the flow rate was incremented manually (note, the pump was also brought into the control loop, and the flow rate could be adjusted using the capacitive sensor signals, see Fig.~S3). All flow rates reported are the measured values calculated by the MCU using the capacitive sensor signals. It was observed that the triggering system robustly separated aqueous from oil signals. The SNR of the TSP signal was determined after Fourier transformation of the FID, whose dependence on flow rate for various sample volumes is plotted in Fig.~\ref{fig:flowandvolume}b and selected spectra are plotted in Fig.~\ref{fig:spectrum}. In general, the SNR decreases as a function of flow rate, as expected, since the sample residence time in the NMR detector also decreases as flow rate is increased. Importantly, the SNR was always enhanced using the automated triggering system as compared to manual triggering; approximately a factor of 2.5, 3.0, and 4.0 improvement in SNR was observed for sample volumes of \SI{120}{\micro\liter}, \SI{80}{\micro\liter}, and \SI{40}{\micro\liter}, regardless of the flow velocity. Conversion of SNR to nLOD\textsubscript{$\omega$} accounts for the differences in sample residence time,\cite{badilita2012microscale} and is plotted in the inset of Fig.~\ref{fig:flowandvolume}b (nLOD\textsubscript{$\omega$} at each flow velocity is plotted in Fig.~S4a). The performance under automated triggering is comparable to previous micro-saddle coil results~\cite{wang2017fast}, with a value of 4-\SI{5}{\nano\mole \ \second^{1/2}}. Under manual triggering conditions, the nLOD\textsubscript{$\omega$} was observed to increase as the sample volume decreased, resulting from the increased contribution of the oil signal (essentially noise) to the FID.

Figure~\ref{fig:flowandvolume}c and d report the results of the second series of experiments, where plugs volume were varied. As in Fig.~\ref{fig:flowandvolume}a, the oil-sample FID signals are recorded separately using the triggering scheme. Seven oil-sample FID pairs are plotted in Fig.~\ref{fig:flowandvolume}c; the sample volume was incremented while the oil volume was held constant at \SI{80}{\micro\liter}. Analysis of the SNR of TSP revealed the expected increase as a function of the sample volume (Fig.~\ref{fig:flowandvolume}d, selected spectra are plotted in Fig.~\ref{fig:spectrum}). Once again, the triggering system yielded a systematic SNR enhancement compared to the manual triggering case, approximately factors of 2.5, 3.0, and 4.0 for sample velocities of \SI{4}{\milli\meter\per\second}, \SI{8}{\milli\meter\per\second}, and \SI{12}{\milli\meter\per\second} regardless of sample volume. The nLOD\textsubscript{$\omega$} for each flow rate (inset, Fig.~\ref{fig:flowandvolume}d) further confirms the performance of both the micro-NMR detector, as well as the enhanced sensitivity when using the automated triggering system (nLOD\textsubscript{$\omega$} calculated at each sample volume is plotted in Fig.~S4b).

\subsection{NMR performance - concentration limit of detection}

\begin{figure*}[t!]
\centering
\includegraphics[width=\textwidth]{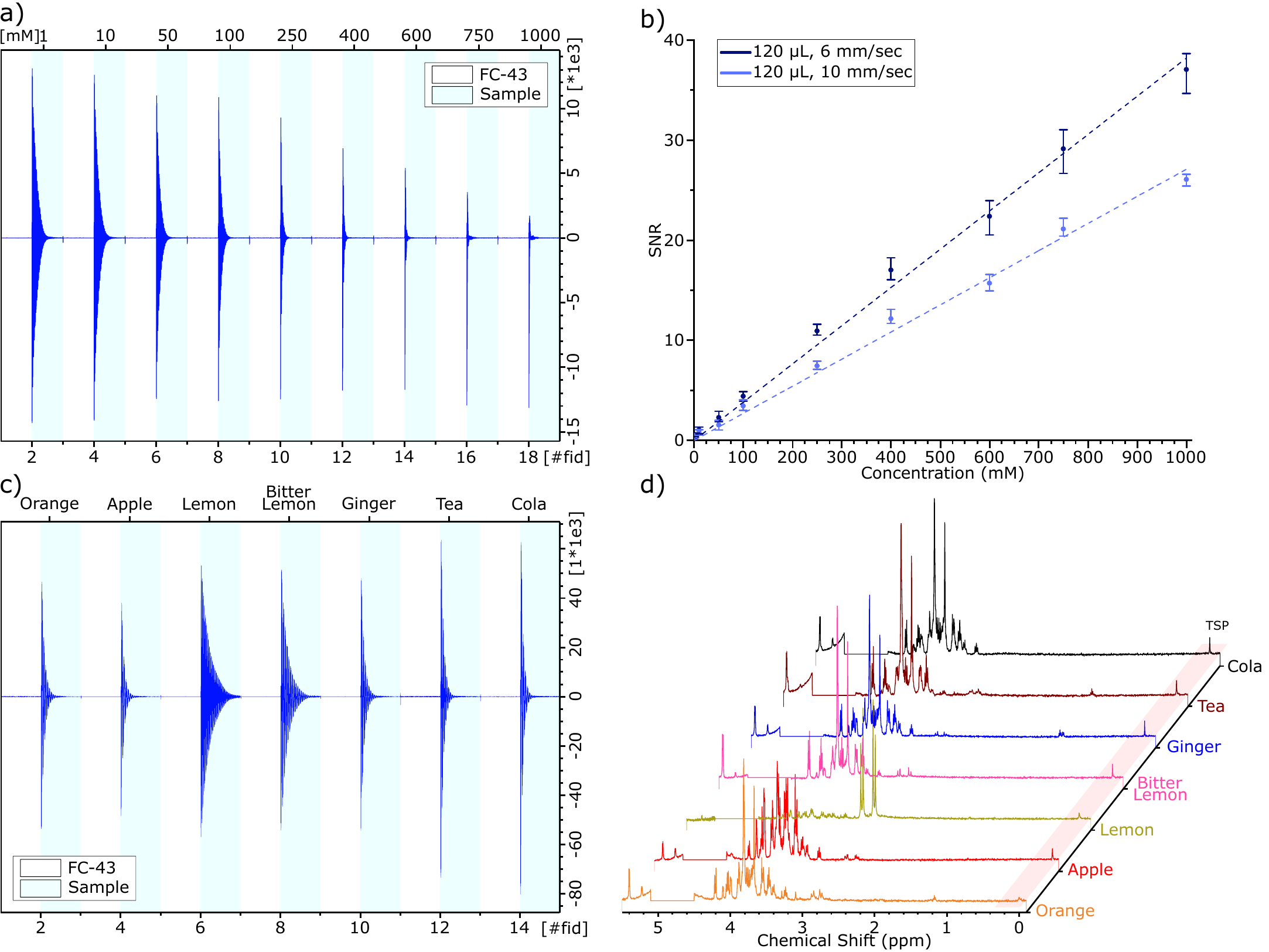}
\caption{(a) Individual FID signals automatically measured from nine aqueous and oil plugs. The aqueous samples were \SI{120}{\micro\liter}, separated by \SI{80}{\micro\liter} oil plugs, and contained glucose concentrations ranging from \SI{0.001}{\molar} to \SI{1}{\molar}. The flow velocity was \SI{10}{\milli\meter\per\second}. (b) SNR versus concentration under two flow velocities. The SNR was calculated using the anomeric hydrogen signal of glucose (5.3-5.1 ppm). Each point is the mean $\pm$ standard deviation of the experiment performed in quadruplicate. Linear fits to the data (dashed lines) had R$^2 >$ 0.99 in both cases. (c) Individual FID signals automatically measured from seven oil and beverage plugs. The oil volume was \SI{80}{\micro\liter} and the sample volume was \SI{120}{\micro\liter}, with a flow velocity of \SI{4}{\milli\meter\per\second}. (d) Resulting NMR spectra from the beverage triggered flow NMR experiment.}
\label{fig:differentsamples}
\end{figure*}

To estimate the dependence on sample concentration using the automated plug-flow based NMR spectroscopy system, the flow tube was loaded with nine aqueous glucose samples spanning a concentration range from \SI{1}{\milli\molar} to \SI{1}{\molar}. All samples plugs were \SI{120}{\micro\liter} and were separated by an FC-43 plug of \SI{80}{\micro\liter}. The automated flow NMR spectroscopy was done at two velocities, \SI{6}{\milli\meter\per\second} and \SI{10}{\milli\meter\per\second}, with the results summarized in Fig.~\ref{fig:differentsamples}a and b. As with previous experiments, the triggering system could robustly separate signals from the oil and aqueous phases (Fig.~\ref{fig:differentsamples}a). The dependence of SNR versus glucose concentration is plotted in Fig.~\ref{fig:differentsamples}b for the two flow rates tested. Here the SNR was calculated by integrating the glucose hydrogen anomeric hydrogen signal (5.1-5.3~ppm).  The SNR was observed to be linear in concentration (R$^2 >$ 0.99) and was greater for the 6 versus \SI{10}{\milli\meter\per\second} flow velocity, essentially translating to the sample having a longer residence time in the NMR detection volume permitting greater signal accumulation. The total experimental time required to measure all oil and aqueous plugs (18 measurements in total) was \SI{6.1}{\minute} and \SI{3.6}{\minute}, or effectively \SI{25.5}{\second} and \SI{15.3}{\second} per sample at flow rates of \SI{6}{\milli\meter\per\second} and \SI{10}{\milli\meter\per\second}. The glucose spectra for these experiments are plotted in Fig.~S6.

\subsection{Automated sample throughput - beverage example}

To validate the system's functionality on a sample set beyond simple test solutions, the tube was loaded with seven samples of different beverages. Each sample plug had a volume of \SI{120}{\micro\liter}, was separated with FC-43 plugs of \SI{80}{\micro\liter}, and the flow rate was \SI{4}{\milli\meter\per\second}. The total NMR acquisition time under these conditions was \SI{7}{\minute} with \SI{38.3}{\second} acquisition time per sample. The resulting FID and spectra of all samples are plotted in Fig.~\ref{fig:differentsamples}c, d. In these samples, TSP was added for a target concentration of \SI{1.25}{\milli\molar}; unfortunately it was observed that the solubility was sample-dependent and therefore the TSP signal could not be used for quantification (only for chemical shift referencing).  As was observed in all previous experiments, i) the automatic triggering system could robustly separate the oil and sample signals; ii) the sample signals were of good spectral quality (both in terms of SNR and resolution); iii) there were no additional signals observed in the oil spectra, ruling out possible diffusion from the aqueous samples (at the NMR detection limit).

\section{Discussion}
The RF coil is the core component of the RF transceiver, enabling the interaction between the measurement device and the sample magnetization. During an MR experiment, the sample experiences a primary static polarizing magnetic field B\textsubscript{0} and, during excitation, a time-varying magnetic field B\textsubscript{1} produced by the RF coil. After excitation, the sample induces an electrical signal in the RF detector as the sample magnetization returns to equilibrium. Here 'coil' is taken to indicate any device capable of producing and detecting magnetic fields, with a variety of geometries satisfying this criterion: 
planar~\cite{Swyer_Interfacing_2016, Montinaro_3D_2018, Swyer_Digital_2019}, 
solenoid~\cite{Li_Multiple_1999, Meier_Microfluidic_2014}, 
Helmholtz~\cite{Spengler2016,C8LC01259H}, 
stripline~\cite{VanBentum2009,Kalfe_Looking,Yilmaz2016,Eills_High_2019}, and 
saddle are examples that have been adapted to microfluidic flow measurement. The saddle configuration (Fig.~\ref{fig:microcoil}a) is an attractive option for tube-based experiments: the cylindrical geometry maximizes the coil filling factor, the detection volume is free from material interfaces perpendicular to B\textsubscript{0}, and recently a robust method for precise microfabrication has been described~\cite{wang2017fast}. These features have translated into the ability to measure high resolution NMR spectra with sensitivity on par with reported micro-detectors with similar detection volumes~\cite{badilita2012microscale}. Spectral resolution, quantified by resonance line width, was found to be \SI{1.25}{\hertz} for the water signal under stationary conditions. The J-coupling of the anomeric glucose hydrogen signal could be readily resolved to $\sim$~\SI{50}{\percent} from baseline resolved (Fig.~\ref{fig:microcoil}e). Sensitivity at \SI{500}{\mega\hertz} was \SI{1.21}{\nano\mol \ \second^{1/2}}, which when scaled to \SI{600}{\mega\hertz} (facilitating comparison with micro-coil reports), converts to \SI{0.88}{\nano\mol \ \second^{1/2}}. For context, Finch \textit{et al}.\ demonstrated nLOD\textsubscript{$\omega$}(\SI{600}{\mega\hertz}) \SI{1.57}{\nano\mol \ \second^{1/2}} sensitivity and $\sim$~\SI{85}{\percent} from baseline resolution of the same signal using their optimized stripline micro-detector~\cite{Finch_An_2016}. Even with the additional noise contribution when connected the active flow-sensing elements (yielding nLOD\textsubscript{$\omega$}(\SI{600}{\mega\hertz}) \SI{1.46}{\nano\mol \ \second^{1/2}}), the NMR sensor performance is acceptable.

Flow sensing based on capacitance measurement has several attractive features. The sensing elements are compatible with standard microfabrication techniques, simplifying the integration with additional functional elements, including NMR spectroscopy as described in this report. The measurement is also non-invasive (avoiding sample contamination, Fig.~\ref{fig:sensor}c) and requires low power (minimizing temperature fluctuations).  As a sensing principle, the limitation is therefore the requirement that individual samples possess a difference dielectric constant. With the device described here, most common solvents can be easily distinguished (Fig.~\ref{fig:sensor}e). By close examination, the two sensors had slightly different responses (Fig.~\ref{fig:sensor}d). The measured capacitance difference between water and FC-43 was $\sim$~\SI{190}{\femto\farad} for the 1\textsuperscript{st} sensor (closest to the CDC chip) and $\sim$~\SI{120}{\femto\farad} for the 2\textsuperscript{nd} sensor. This difference between the two sensors can be attributed to the parasitic capacitance built-up by the long tracks. Nevertheless, the performance was comparable to other systems~\cite{elbuken2011detection, zargar2018novel} implementing that same sensing concept, even with the effect of the parasitic capacitance. A sensitivity advantage was realized by rolling the sensors around the flow types, enhancing the sensitivity to the change in the dielectric constant compared to planar IDC sensors~\cite{elbuken2011detection} and sensors with less filling factor~\cite{zargar2018novel}. The sensor's sensitivity limit, excluding the effect of parasitic capacitance, was calculated to be $\Delta\epsilon_r= 0.2$.

Optical methods could also be used for non-invasive sensing, extending the flexibility in sample handling. Optical techniques would require waveguides to pass the light through the sample, and either optical density of spectrometer detection systems to monitor differences in scattering or chemical composition as a means to track individual sample positions. Regardless of the source of the sample-specific signal, once digitized the trigger processing pipeline described in this report could be easily recruited for NMR synchronization. 

NMR spectroscopy is characterized as high resolution, with spectral line widths on a p.p.b. scale. This is the result of nuclear spins possessing relatively long relaxation times, often on the order of seconds for hydrogen nuclei in liquid state.  One cost to be paid for this is the relatively long re-equilibration times, thus limiting how often an experiment can be repeated. Under flow conditions, this recycle delay can be minimized since freshly polarized sample is constantly transported into the detection volume, removing the dependence on re-equilibration. This is especially important in cases where nuclei have long relaxation time constants~\cite{Carret_Inductive_2018}. Combining the long experiment recycle times with the fact NMR is relatively insensitive results in the reality that NMR is a low throughput method. Sample exchange is not always the time limiting step, and therefore automated sample exchange systems based on individual glass NMR sample tubes have been adequate and even the standard. Commercial systems are additionally available that will load the glass NMR tubes. In conditions in which sample exchange does become limiting, a flow-based approach has the ability to outperform the standard sample handling strategy and increase throughput. There are several experimental scenarios that fall into this category: food and beverage screening, biofluid metabolomics, and chemical library screening are examples.  If we consider a PTFE tube with \SI{1.0}{\milli\meter} I.D., \SI{100}{\micro\liter} aqueous and \SI{100}{\micro\liter} oil plugs, then in \SI{100}{\meter} of this PTFE tube nearly 400 individual samples could be loaded (x2 if the immiscible phase also carries measurable molecules). Loading the PTFE is the time limiting step, although this could also be automated (under investigation in our group). Under measurement automation, a reasonable flow rate (e.g. \SI{4}{\milli\meter\per\second}), assuming an acquisition time of \SI{40}{\second} per sample (as was used in the beverage example), and both phases contain measurable content, then these 800 samples would require $\sim$~\SI{9}{\hour} of measurement time. Note that if the immiscible phase acts only to separate samples, then 400 samples would also require $\sim$~\SI{9}{\hour}.  This could be improved by bringing the pump into the control loop and varying the flow velocity, so that the oil phase passes through the detection volume at an increased rate.  As comparison, 300 samples required \SI{120}{\hour} of measurement when done manually (\SI{500}{\micro\liter} per sample, \SI{5}{\milli\meter} sample tubes, \SI{7}{\minute} acquisition time / sample neglecting time for instrument adjustment after each sample exchange, without automatic sample exchange system)~\cite{MacKinnon_NMR_2019}, suggesting over an order of magnitude acceleration is possible. 

To achieve enhanced throughput, a balance must be found between flow rate, SNR, and spectral resolution. As demonstrated in Fig.~\ref{fig:spectrum}, increasing the flow rate leads to a degradation of SNR and spectral resolution, all other parameters held constant. As the flow rate is increased, the NMR measurement time is concomitantly reduced and thus there is less noise averaging over the accumulated signals. At increased flow rates, a larger fraction of the excited sample leaves the NMR detection volume during signal acquisition, resulting in an artificial shortening of the FID and manifesting as an increase in the effective relaxation time T\textsubscript{2}. This results in spectral line broadening, as observed in Fig.~\ref{fig:spectrum}. Based on the experimental requirements, the flow velocity can be adjusted: high concentration, single molecule samples could be measured at high flow velocity for identification purposes (sacrificing J-coupling resolution); low-concentration, complex mixtures will require lower flow velocity to enable signal accumulation and maximum spectral resolution. 

\section{Materials and Methods}
\subsection{Triggering system}
\begin{figure}[t!]
\centering
\includegraphics[width=0.5\columnwidth]{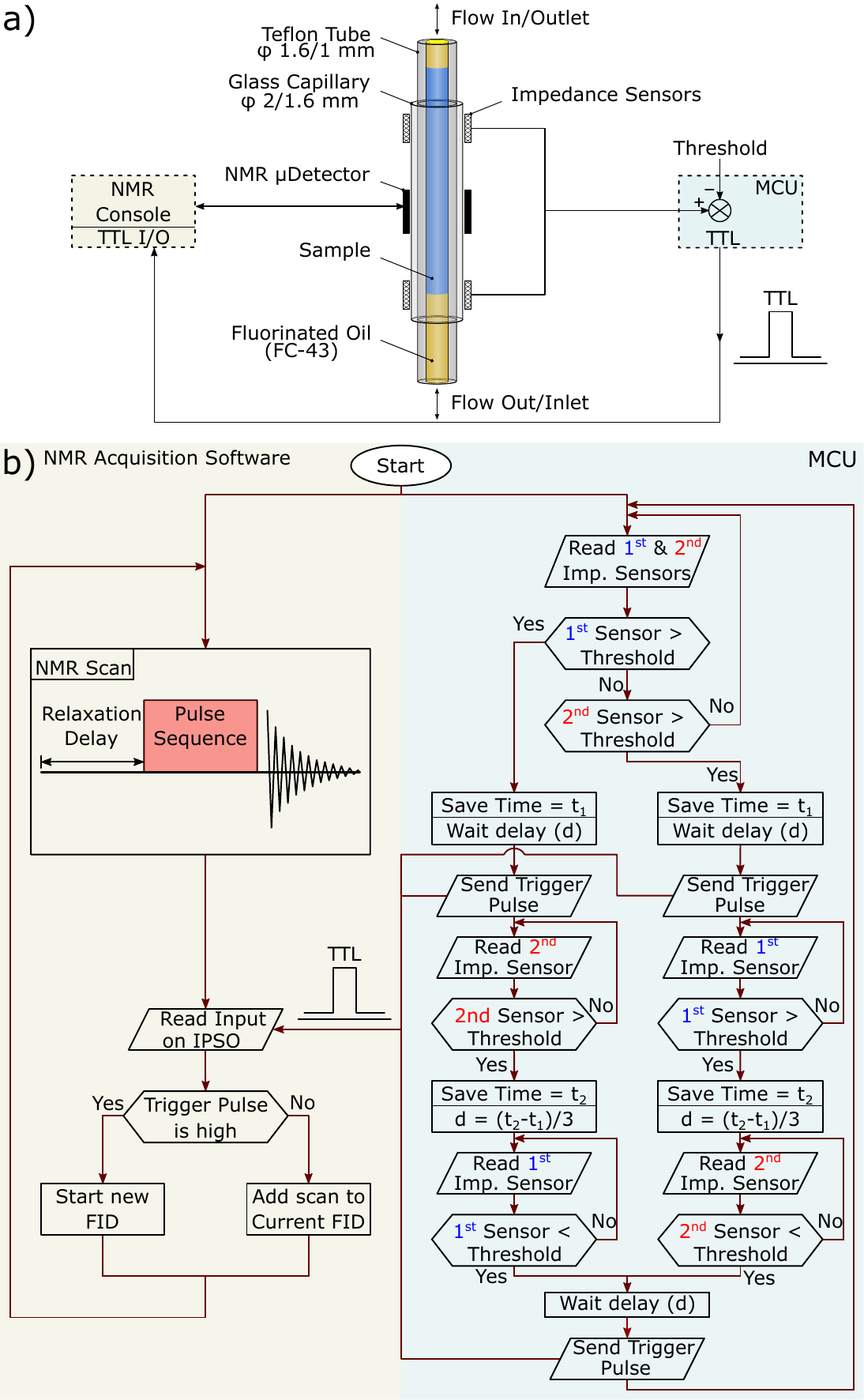}
\caption{(a) A schematic diagram illustrating the main components of the control system. The microcontroller unit (MCU) synchronizes the NMR experiment by sending the TTL signal to the spectrometer console's TTL I/O unit. The MCU triggers the TTL pulse based on the capacitive sensors value. (b) A flow diagram describing the executed algorithms on both the NMR acquisition software and the MCU. The two algorithms execute in parallel, where the MCU's algorithm triggers the NMR pulse sequence to increment the FID based on the presence of a sample.}
\label{fig:algorithm}
\end{figure}

A schematic diagram of the control algorithm is shown in Fig.~\ref{fig:algorithm}a. During the flow, a change in capacitance of the sensor electrodes is detected when the effective dielectric interacting with the fringe electric field changes. The changes in capacitance are then sensed by the microcontroller unit (MCU), which in turn triggers the NMR experiment.

During the synchronized NMR experiment, two software algorithms are running in parallel, one on the MCU and the other is the pulse sequence on the NMR acquisition software (TopSpin 3.5pl2, Bruker BioSpin). A flow diagram for the two algorithms is presented in Fig.~\ref{fig:algorithm}b. Following the system start, the NMR acquisition software starts collecting the NMR scans of the first FID. After each scan, the NMR software checks the input value on the TTL signal pin in the TTL I/O unit. In case the TTL signal is low, the spectrometer adds the current scan to the same FID. If the TTL signal is high, the spectrometer saves the current FID to disk after adding the last scan and then initiates a new FID measurement. Meanwhile, the program on the MCU monitors the capacitance of the two sensors. Once the capacitance measurement of either of the two sensors exceeds a user-set threshold value, the time is recorded ($t_1$), and a trigger signal is sent from the MCU to the TTL I/O unit of the NMR console after a delay ($d$). Afterward, the MCU monitors the value of the other sensor until it exceeds the threshold value and then records the time ($t_2$). Subsequently, the MCU monitors the first sensor until the value becomes lower than the threshold value, after which a trigger signal is sent after a delay ($d$). Hence, for a single sample plug, the TTL I/O unit receives two TTL pulses, the first pulse when a sample enters the NMR detector (start new FID acquisition) and the second pulse when it leaves the detector (save current FID). This control algorithm results in saving all the scans collected for individual samples in a single FID. Afterward, the MUC checks again in a threshold detection loop for both sensors' capacitance, waiting for the next sample. The MCU checks both sensors at the beginning to maintain the flexibility of bidirectional flow.

The TTL signal is always sent to the TTL I/O unit after a delay ($d$) to ensure that the sample either reached or left the NMR detector. The delay ($d$) is calculated using Equation~\ref{eq:delay}, taking into account the distance between the sensor and the NMR detector \SI{10}{\milli\meter}, the gap between the two sensors \SI{30}{\milli\meter}, and the recorded times ($t_1$ and $t_2$). The initial value of $d$ is set to zero and updated after measuring the velocity of each plug. Hence, a dummy plug is used at the beginning to determine the first value of the delay.
\begin{equation}
d = \frac{t_2-t_1}{3}.
\label{eq:delay}
\end{equation}

To ensure that the NMR acquisition software captures the TTL pulse, the signal width (t$_{TTL}$) should be longer than the sum of the NMR scan acquisition time ($T_{AQ}$) and relaxation delay ($D1$). The TTL pulse must also be shorter than twice the acquisition time + relaxation delay; otherwise, it might be captured twice by the NMR software:
\begin{equation}
(T_{AQ}+D1) < \textbf{t$_{TTL}$} < 2\times(T_{AQ}+D1).
\label{eq:TTLpulse}
\end{equation}

\subsection{Microfabrication}
\begin{figure*}[t!]
\centering
\includegraphics[width=\columnwidth]{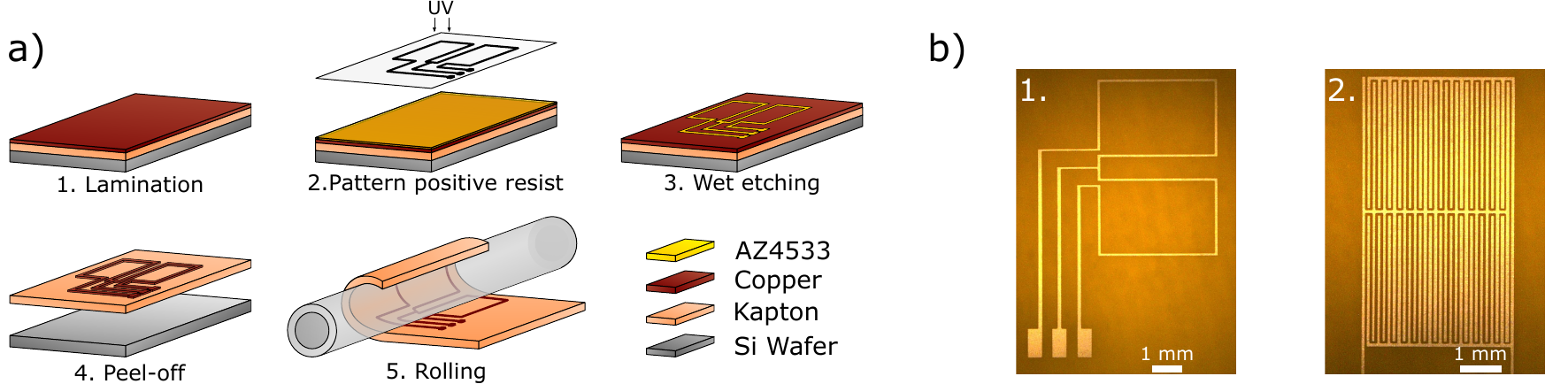}
\caption{(a) Schematic illustration of the microfabrication process. First, a wet etching photolithography process: (1) substrate preparation, by laminating Kapton film covered with a conductive layer on a silicon wafer. (2) Positive photoresist (AZ4533) patterning using UV lithography. (3) etching the exposed regions of the conductive layer, and (4) peeling off the film from the wafer. Second process: (5) rolling the patterned conductive layer on the Kapton film around the glass capillary. (b) Microscope images of the 2D patterned (1) saddle coil and (2) capacitive sensor.}
\label{fig:fabrication}
\end{figure*}

Fabrication of the micro saddle detector and the interdigitated capacitive sensors around a slender glass capillary with outer diameter \SI{2.15}{\milli\meter} was done in two main steps, as shown in Fig.~\ref{fig:fabrication}a. The first step was 2D patterning of a conductive layer on a flexible polymer film in the form of the coil and the sensor. The second step was rolling the structured film around the capillary tube as described in~\cite{wang2017fast}, thereby forming a 3D structure around the glass capillary. 2D patterning of the conductive layer was done following a standard photolithography process, which is suitable for mass production at the wafer scale. A \SI{25}{\micro\meter} Kapton film covered from one side with a \SI{9}{\micro\meter} copper layer (AkaFlex KCL 2-9/25 HT, Krempel GmbH) was used as the base substrate for the photolithography process. Kapton polymer was chosen as it has a magnetic susceptibility close to that of water~\cite{wapler2014magnetic}. With a thickness of \SI{25}{\micro\meter} the film is flexible enough to be rolled around the glass capillary and at the same time, it can be mechanically handled during the fabrication process. A copper layer of \SI{9}{\micro\meter} provides the required skin depth of the conductor at \SI{500}{\mega\hertz}, ensuring the minimal resistive losses of the NMR detector. To facilitate the polyimide film handling during the fabrication process, it was cut using an infrared laser cutter (VLS3.50, Universal Laser System) into the shape of \US{4}{\inch} wafer, then laminated to a silicon wafer via a \SI{3.3}{\micro\meter} AN AZ4533 photoresist layer (Micro-Chemicals GmbH), spin-coated at \SI{4000}{\rpm}. After lamination, the wafer was soft-backed at \SI{95}{\degree} for \SI{1}{\min}. The wafer was dipped into a sodium persulfate solution(\ch{Na2S2O8}, \SI{240}{\gram\per\liter}) for \SI{15}{\sec} to clean the film by etching the copper surface; also, this step promotes photoresist adhesion. Afterward, the wafer was cleaned with DI water and dried with a pressurized nitrogen stream, directly after that a \SI{3.3}{\micro\meter} layer of AZ4533 positive photoresist was spin-coated on top of the film to prevent oxidation. The wafer was then soft-baked at \SI{95}{\degree} for \SI{1}{\min}. The photoresist was structured by UV exposure at a dose of \SI{235}{\milli\joule\per\square\centi\meter}. A KOH based AZ400K developer, prepared at 1:4 dilution, was used for developing the photoresist for \SI{3}{\min}. After cleaning with DI water, the wafer was immersed in a sodium persulfate solution for etching the exposed copper. The film was peeled off from the silicon wafer and clean from the backside with acetone. The film was fixed on a glass capillary (OD/ID, \SI{2.15/1.72}{\milli\meter}, borosilicate glass, Hilgenberg GmbH) using epoxy (UHU Plus). The glass capillary was glued to the PCB using the same epoxy. A PCB holder was 3D printed from polylactide polymer (PLA) using an Ultimaker~2\textsuperscript{+} printer.

\subsection{Coil characterization and interfacing}
With a rolled-up saddle coil connected to a custom-built probe, pre-matching and tuning was performed by modifying non-magnetic capacitors (S111DUE, Johanson Technology) soldered on the PCB and validated using a network analyzer (Agilent, E5071B). The pre-tuning and matching circuit on the probe head consist of a symmetric configuration of C\textsubscript{M,PCB}-(L$\parallel$C\textsubscript{T,PCB})-C\textsubscript{M,PCB}, where C\textsubscript{T,PCB} is used to tune the coil to the \textsuperscript{1}H NMR frequency, while matching to \SI{50}{\ohm} is realized with C\textsubscript{M,PCB} capacitors which are connected to the \textsuperscript{1}H channel connectors of the probe. The custom-built probe posses externally controllable matching and tuning trimmers for fine-tuning (\SI{0.5}{} - \SI{10}{\pico\farad}, TG 092 ROHS, EXXELIA). The reflection coefficient S\textsubscript{11} was used to check the resonance. The coil was consecutively tuned to \SI{500}{\mega\hertz}, which is the \textsuperscript{1}H proton resonance frequency (Larmor frequency) at \SI{11.7}{\tesla}, and matched to \SI{50}{\ohm} as shown in Fig.~\ref{fig:microcoil}c. The quality factor was calculated with Equation~\ref{eq:qfactor} knowing the bandwidth at \SI{-3}{\decibel} ($f\textsubscript{\SI{-3}{\decibel}}$) of the resonance frequency $f\textsubscript{resonance}$,
\begin{equation}
    Q = \frac{2f\textsubscript{resonance}}{\Delta f\textsubscript{\SI{-3}{\decibel}}}.
    \label{eq:qfactor}
\end{equation}

\subsection{Materials}
D-glucose, 3-(trimethylsilyl)- propionic-2,2,3,3-d4 acid sodium salt (TSP), and deuterium oxide  (D\textsubscript{2}O, 99.9\% deuterium, Sigma Aldrich) were purchased from Sigma-Aldrich. A market available coloring power (Brauns Heitmann) was used. The Fluorinert oil FC-43 ( $>99\%$, 3M\textsuperscript{TM} Fluorinert Electronic Liquid) was purchased from Ionic Liquids Technologies (IOLITEC) GmbH. A selection of commercial beverages were privately donated.

\subsection{Simulation}

COMSOL Multiphysics software was used for electromagnetic simulations of the micro saddle coils and the capacitive sensors.

The micro saddle coil geometry was created around two concentric tubes, to model a glass capillary and a Teflon tube filled with water. The saddle coil with the tubes was located inside a cylinder-shaped domain to model the surrounding air. Two simulation scenarios were considered:
\begin{enumerate}[label=\Roman*]
    \item The saddle coil was connected to a \SI{50}{\ohm} port for excitation by \SI{1}{\ampere} signal at \SI{500}{\mega\hertz}. In this model, the COMSOL radio-frequency module (electromagnetic waves, frequency domain) was used. A radiation boundary condition was added, which acted as an absorbing layer to prevent back-scattering. The generated B\textsubscript{1} field was deduced and the coil sensitivity $\eta\textsubscript{RF}$ was calculated with Equation~\ref{eq:coilsensitivity}~\cite{badilita2012microscale}, taking into account the coil's high-frequency resistance $R$ and the electrical driving power $P$,
    \begin{equation}
    \eta\textsubscript{RF} = \frac{B\textsubscript{1}}{\sqrt{P}} = \frac{B\textsubscript{1}}{i\sqrt{R}}.
    \label{eq:coilsensitivity}
    \end{equation}
    
    \item An external static magnetic field of \SI{11.7}{\tesla} was applied along the axis of the tubes. In this model, the COMSOL AC/DC module (magnetic fields, no currents) was used. This simulation model shows the effect of the coils tracks on the static B\textsubscript{0} field.
\end{enumerate}

In both simulation scenarios, the gap between the two rectangular loops of the saddle coil ($\theta\textsubscript{2}$) was swept to obtain the highest sensitivity with the most homogeneous B\textsubscript{0} and B\textsubscript{1} fields. The gap was swept so that the gap perimeter ($\theta\textsubscript{2}\times r$) changed from \SI{100}{\micro\meter} to \SI{1300}{\micro\meter} with a step of \SI{100}{\micro\meter}. 

To simulate the electric field of the capacitive sensor, the sensor was constructed around a glass and Teflon tubes placed in an air like environment. The domain inside the Teflon tube was configured, using the relevant parameters, once with for water and again for FC-43. The sensor was excited with a sinusoidal voltage of \SI{5}{\volt} at \SI{32}{\kilo\hertz} (simulating the parameters used in the existing system), and the electric field was plotted.

\subsection{Sample preparation and NMR spectroscopy}

The test sample used for evaluating the SNR of the flow system was prepared as \SI{300}{\milli\molar} glucose, \SI{75}{\gram\per\liter} coloring powder, and \SI{30}{\milli\molar} TSP dissolved in deionized (DI) water. 
The glucose concentration series from \SI{1}{\milli\molar} to \SI{1}{\molar} was prepared in using 50:50 DI water:D\textsubscript{2}O. 
Beverage samples were prepared by adding \SI{1}{\milli\liter} of \SI{5}{\milli\molar} TSP in D\textsubscript{2}O to \SI{3}{\milli\liter} of the beverage  ([TSP]$_{final}$~=~\SI{1.25}{\milli\molar}). 

The microfluidic tube used for sample flow was polytetrafluoroethylene (PTFE) (OD/ID, \SI{1.6/1}{\milli\meter}, S1810-12, Bohlender GmbH) and had a length of \SI{4.5}{\meter}.
After feeding the Teflon tube through the probe and the capillary tube on the probe head, one terminal of the tube was connected to a syringe pump while the other terminal was left open. Initially, the syringe and the whole line were filled with FC-43 to avoid air gaps, which hinder the stability of the flow. Then the syringe pump was adjusted to the required volume, and the pump was switched on to draw the sample by keeping the free terminal of the tube in the sample reservoir. Afterward, the tube's terminal was moved to the FC-43 reservoir to draw an oil plug for sample separation before moving to the next sample reservoir. After each filling step, the tube terminal was immersed in a pure DI water reservoir to avoid cross-contamination. During the NMR experiments, the flow was continuously on.

NMR experiments were performed on an \SI{11.7}{\tesla} wide-bore AVANCE III NMR spectrometer ($^1$H resonance frequency \SI{500}{\mega\hertz}).  The pulse sequence was a modified version of a single pulse experiment, including the trigger signal control. For a typical FID, 5998 points were collected over a spectral width of \SI{10}{\kilo\hertz}, using a \SI{0.1}{\second} recycle delay. For the beverage NMR experiments, the number of points was increased to 12014. The FIDs were collected in a pseudo-2D NMR experiment, which were separated and processed individually after the experiment was complete. Prior to Fourier transformation, each FID was zero filled by a factor of 2 and multiplied by an exponential function equivalent to \SI{0.3}{\hertz} line broadening. Each spectrum was referenced to TSP (0 ppm), phase, and baseline corrected.

\section{Conclusion}
The challenge of sample throughput in NMR spectroscopy can be addressed by implementing a plug flow-based approach.  By implementing an independent measurement mode for flow position and velocity, a level of automation is also added to the NMR signal acquisition system, further improving throughput. The flow-sensing approach described in this report is an additional step towards increasing the functionality of the NMR system, enabling increased experimental flexibility and sophistication, or enhanced SNR w.r.t. standard equipment by dividing a large sample into small individually acquired droplets, each with high SNR. Active components near to the NMR sensor do impact sensitivity, and this point must be considered in the system design and can be engineered out. Nevertheless, spectral performance was sufficient for the measurement of commercial beverages, without sample treatment and even after a factor of 1.3 dilution.  While not demonstrated here, the platform is compatible with parallelization as recently described for low-field NMR spectroscopy~\cite{Lei_Portable_2020}, potentially further increasing sample throughput scaled by the number of parallel NMR sensors.

\section*{Acknowledgements}
ON sincerely acknowledges financial support from the DAAD (German Academic Exchange Service) under the German Egyptian Research Long-Term Scholarship Program (GERLS) and the BioInterfaces International Graduate School (BIF-IGS) at the Karlsruhe Institute of Technology (KIT). ON thanks his KIT colleagues Hossein Davoodi, Erwin Fuhrer, and Nurdiana Nordin for the fruitful conversations. DM and JGK acknowledge support from the European Union’s Future and Emerging Technologies Framework (Grant H2020-FETOPEN-1-2016-2017-737043-TISuMR). JGK acknowledges partial support from ERC Synergy Grant HiSCORE 951459. MJ, DM, JGK, and NM acknowledge the partial financial support of the Helmholtz Association through the programmes “Science and Technology of Nanosystems - STN”, and “BioInterfaces in Technology and Medicine - BIFTM". All Authors would like to acknowledge the support of the Karlsruhe Institute of Technology (KIT), providing the infrastructure to realise this work.

\bibliographystyle{naturemag}
\bibliography{main}

\newpage

\date{}                  
\setcounter{Maxaffil}{0}
\renewcommand\Affilfont{\itshape}
\setcounter{section}{0}
\setcounter{table}{0}
\renewcommand{\thetable}{S\arabic{table}}%
\setcounter{figure}{0}
\renewcommand{\thefigure}{S\arabic{figure}}

\setstretch{1.5}

\section*{Supplementary information}

\begin{figure*}[ht]
\centering
\includegraphics[width=\textwidth]{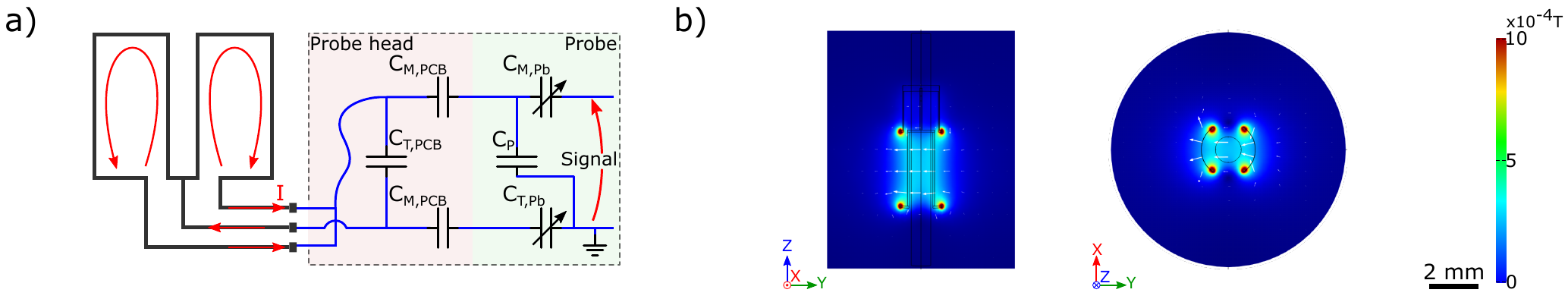}
\caption{a) The electrical connection of the NMR saddle detector. The three-terminals of the saddle detector are connected in a way to keep the current flowing in the opposite direction in each loop. Hence, the two magnetic fields will constructively interfere inside the coil (sample region). b) The simulated magnetic $B_1$ fields of the two loops are pointing in the same direction.}
\label{fig:coil_magnetic_field}
\end{figure*}

\begin{figure*}[ht]
\centering
\includegraphics[width=0.8\textwidth]{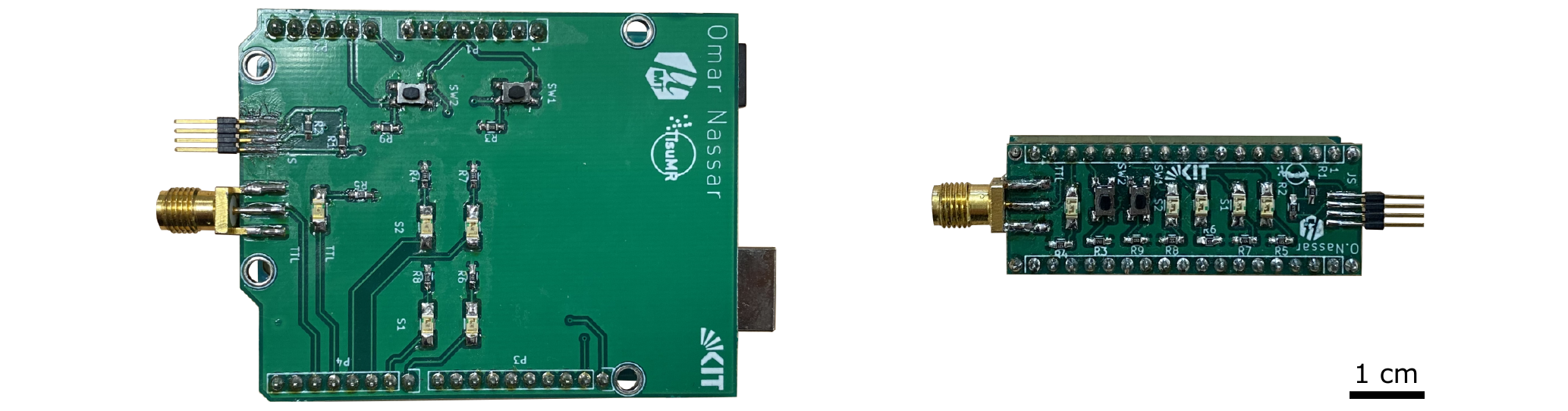}
\caption{Two Arduino shields were fabricated and used for the NMR experiments. One shield compatible with Arduino Uno (left), and the other shield compatible with Arduino micro (right). The performance of the microcontrollers was equivalent.}
\label{fig:arduino_shields}
\end{figure*}

\begin{figure*}[ht]
\centering
\includegraphics[width=\textwidth]{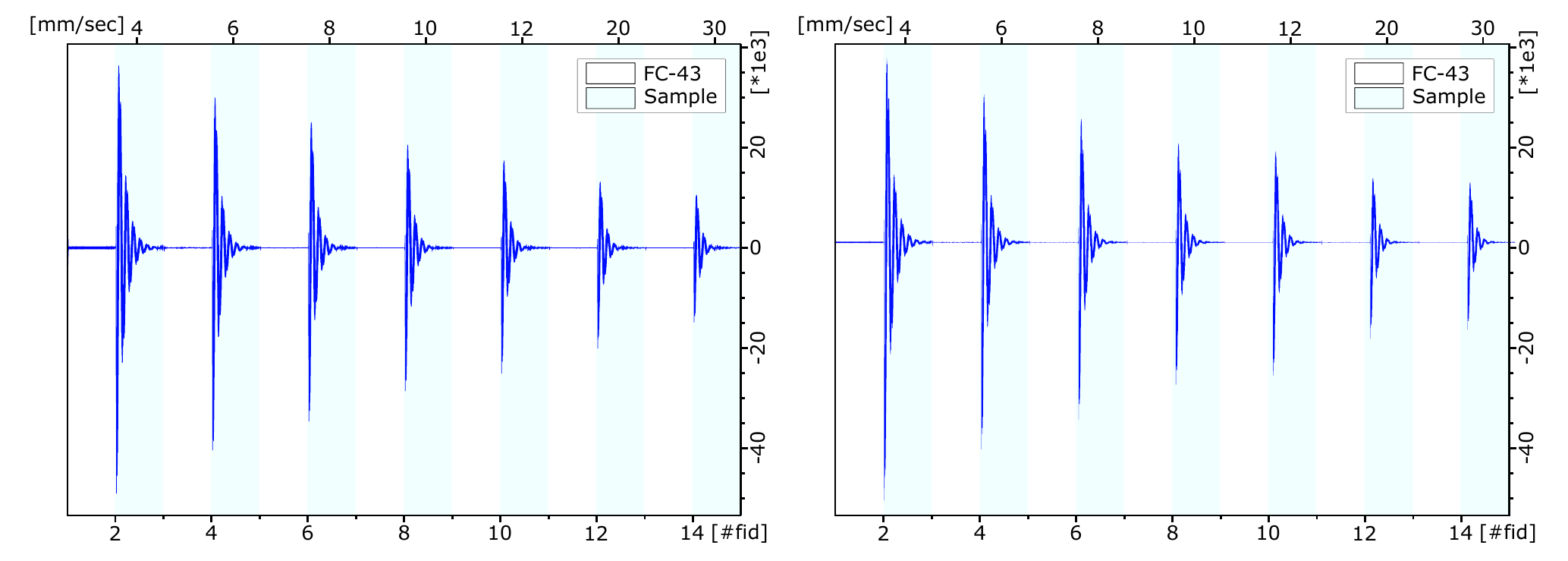}
\caption{Flow rate control while performing the flow NMR experiment can be done manually (right) or automatically (left). In both cases, no noticeable effect on the spectrum was observed. In the automatic control case, the syringe pump (Fusion 100-X Touch,  KR Analytical) was used, controlled via LabVIEW software over a Universal Serial Bus (USB) connection. In this experiment, the capacitance signal was used to update the flow rate in real-time, i.e., after the NMR signal acquisition of an aqueous sample was completed, the same trigger signal sent to the spectrometer used to save the FID was also used to instruct the pump to change the flow rate.}
\label{fig:pumpcontrol}
\end{figure*}

\begin{figure*}[ht]
\centering
\includegraphics[width=\textwidth]{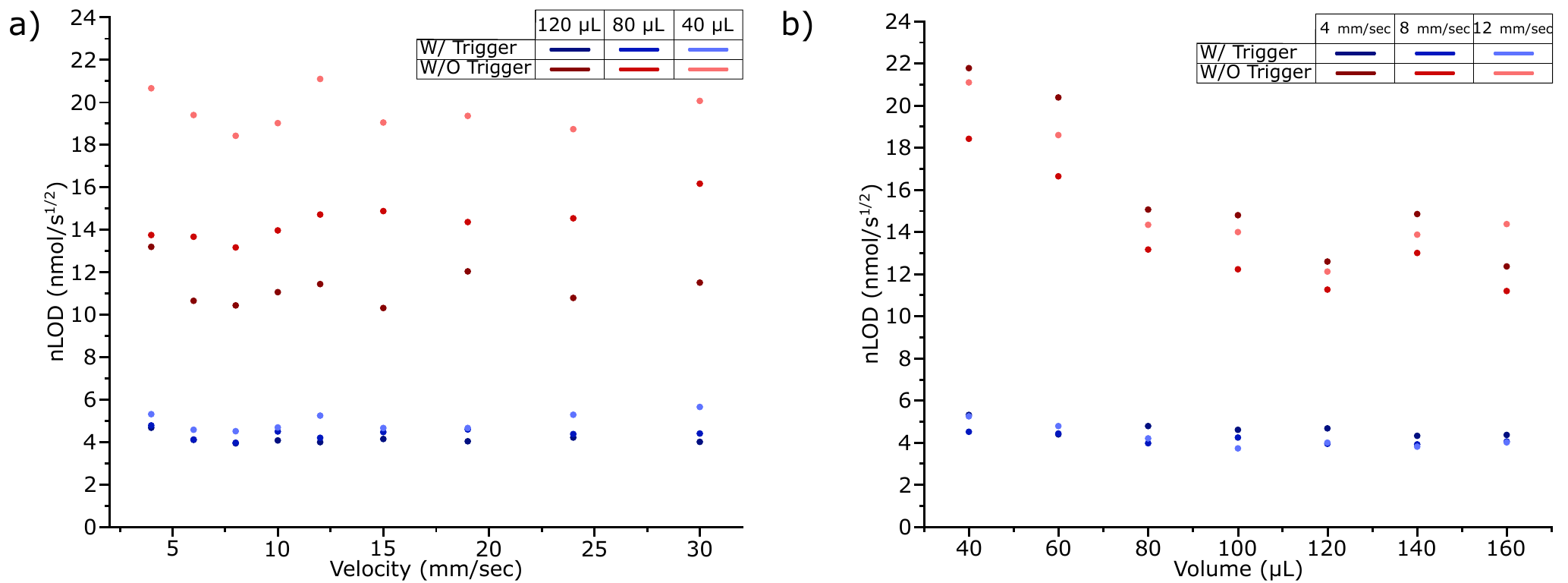}
\caption{a) The normalized limit of detection (nLOD\textsubscript{$\omega$} ) versus sample velocity.  The nLOD\textsubscript{$\omega$} varies within a small range over the enter range of velocities for a given volume.
Under manual triggering conditions, the nLOD\textsubscript{$\omega$} increases as the sample volume decreased due to the increased contribution of the oil signal to the FID (noise). b) The normalized limit of detection (nLOD\textsubscript{$\omega$}) versus sample volume.  Under auto-triggering conditions, a minimal enhancement in the nLOD\textsubscript{$\omega$} is achieved by increasing the sample volume. On the other hand, in manual triggering, the nLOD\textsubscript{$\omega$} is highly enhanced by increasing the sample volume as the signal contribution from the oil plugs (i.e. noise) is reduced.  The mean~$\pm$~standard deviation of these data are plotted in Fig.~4, main text.}
\label{fig:nLOd}
\end{figure*}

\begin{figure*}[ht]
\centering
\includegraphics[width=\textwidth]{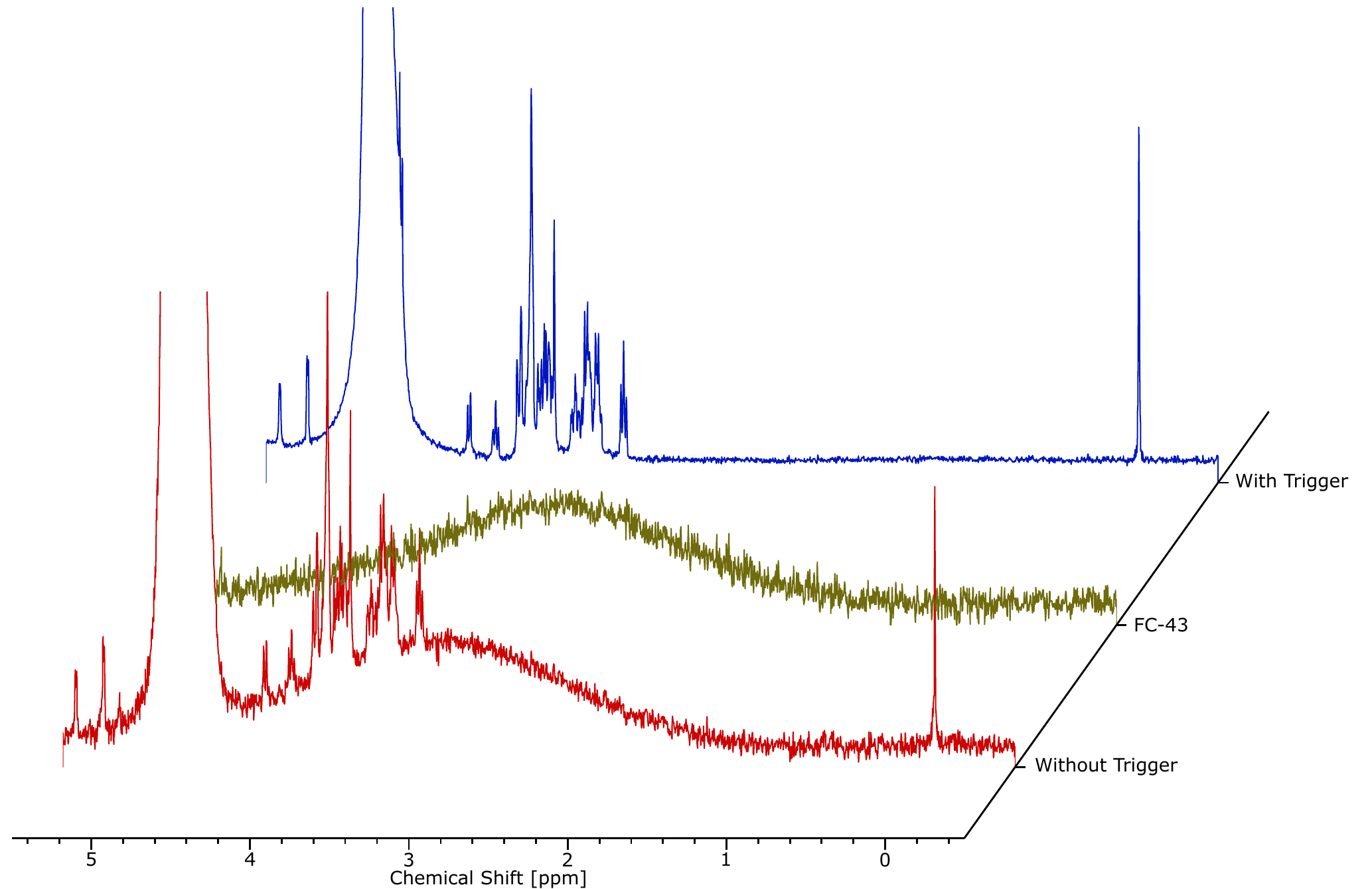}
\caption{Representative $^1$H NMR spectra. The blue spectrum is of a \SI{120}{\micro\liter} sample plug (\SI{300}{\milli\molar} glucose, \SI{75}{\gram\per\liter} coloring powder, and \SI{30}{\milli\molar} TSP dissolved in DI water) acquired using the auto triggering system. In contrast, the red spectrum of the same sample obtained with manual triggering. The flow velocity was \SI{4}{\milli\meter\per\second}.}
\label{fig:withvswithoutTTL}
\end{figure*}

\begin{figure*}[ht]
\centering
\includegraphics[width=\textwidth]{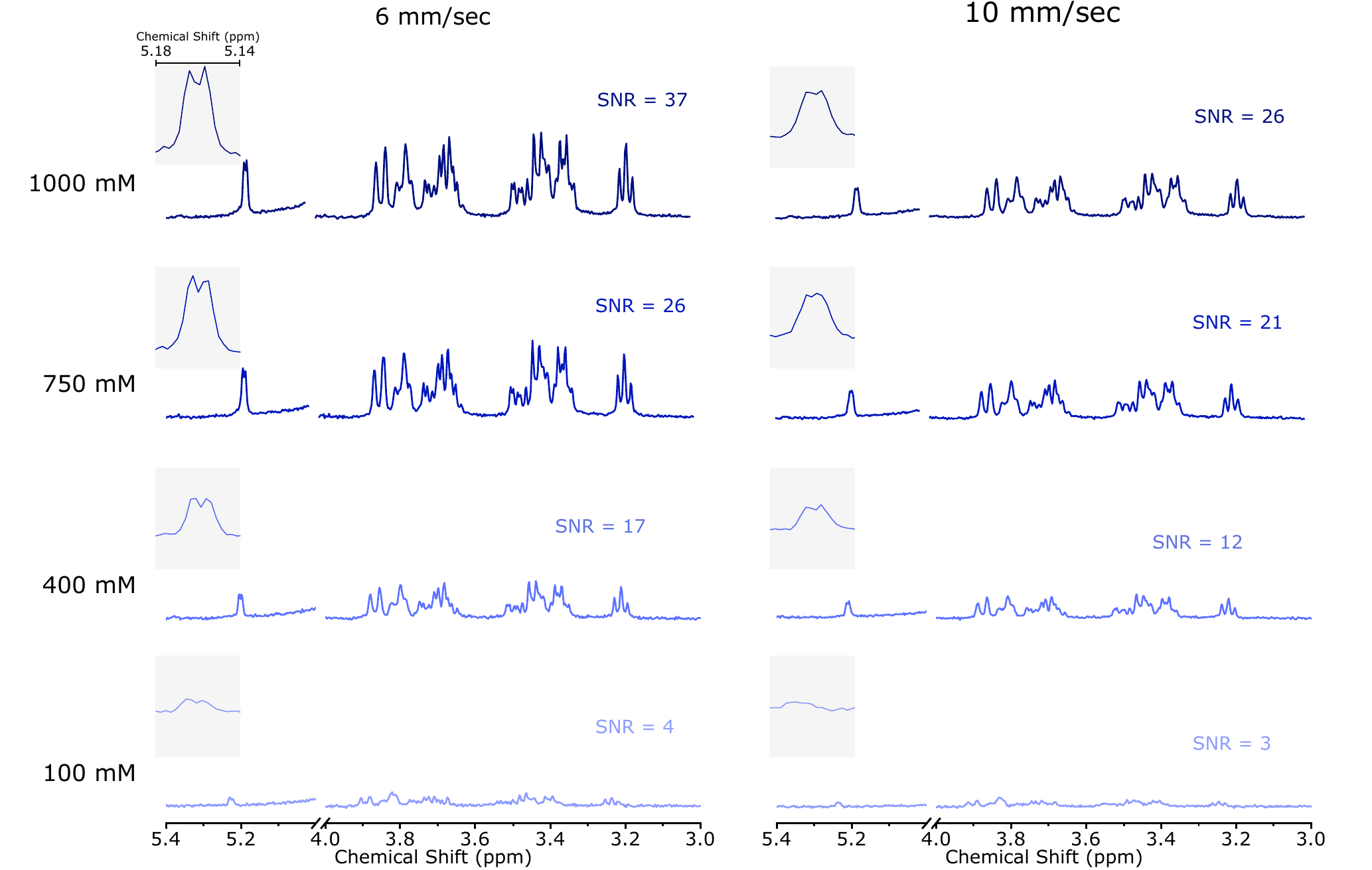}
\caption{Selected $^1$H NMR spectra from the flow NMR spectroscopy experiments varying the glucose concentrations in DI water. The SNR of the glucose anomeric proton for each spectrum is noted next to the signal with an expanded view of the peak (5.18-5.14 ppm) for spectral resolution comparison.  The spectral region between 4-5 ppm (water resonance) was excluded for clarity.}
\label{fig:Concentration}
\end{figure*}

\begin{figure*}[ht]
\centering
\includegraphics[width=0.7\textwidth]{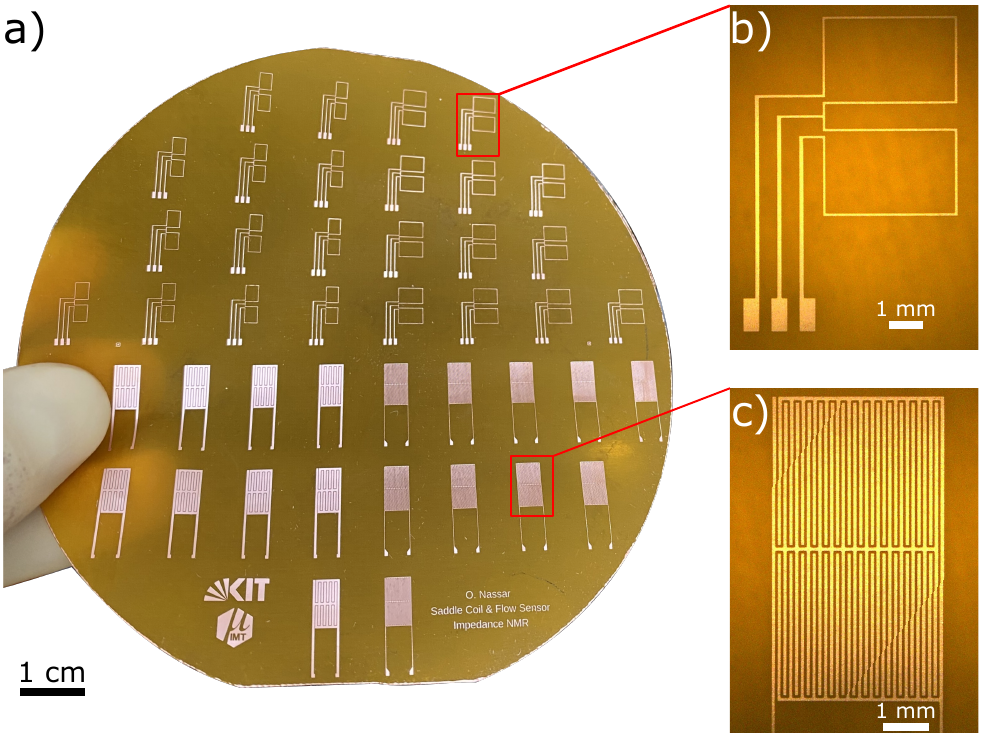}
\caption{a) A photograph of a fully patterned Kapton film in the shape of a \SI{100}{\milli\meter} wafer. The film possesses a batch of the micro saddle NMR detector and the interdigitated capacitive sensor, demonstrating the developed process's compatibility with mass production. The sub-figures show microscopic images of the 2D patterned (b) saddle detector and (c) interdigitated capacitive sensor.}
\label{fig:wafer}
\end{figure*}

\end{document}